\documentclass[aps,10pt,a4paper,twocolumn,english,prl,showpacs,noshowkeys,nosuperscriptaddress]{revtex4-1}

\usepackage{mathpazo}

\usepackage[latin1]{inputenc}
\usepackage{color}
\usepackage{babel}
\usepackage{amsmath}
\usepackage{graphicx}
\usepackage[unicode=true,pdfusetitle,
 bookmarks=true,bookmarksnumbered=false,bookmarksopen=false,
 breaklinks=false,pdfborder={0 0 0},backref=false,colorlinks=true]
 {hyperref}
\usepackage{breakurl}

\makeatletter

\@ifundefined{definecolor}
 {\usepackage{color}}{}
\usepackage{babel}
\hypersetup{colorlinks,citecolor=blue,filecolor=black,linkcolor=red,urlcolor=blue,pdftex}
\usepackage{colortbl}

\makeatother

\begin{document}

\title{Modulation instability in high-order coupled nonlinear Schr\"odinger
equations with saturable nonlinearities}

\author{Erivelton O. Alves}
\affiliation{Instituto de F\'isica, Universidade Federal de Goi\'as, 74.001-970, Goi\^ania,
Goi\'as, Brazil}
\author{Wesley B. Cardoso}
\affiliation{Instituto de F\'isica, Universidade Federal de Goi\'as, 74.001-970, Goi\^ania,
Goi\'as, Brazil}
\author{Ardiley T. Avelar}
\affiliation{Instituto de F\'isica, Universidade Federal de Goi\'as, 74.001-970, Goi\^ania,
Goi\'as, Brazil}

\begin{abstract}
The influence of a saturable nonlinearity on the modulation instability
in oppositely directed coupler in the presence of high-order effects
is investigated. By using the standard linear stability analysis,
we obtain the instability gain that exhibits a significant change
in the bands of instability due to the effects of a saturable nonlinearity.
We also show that even in the presence of saturation there is no change
in instability gain when we compare the results obtained for both
channels not influenced by self-steepening effect or for both channels
influenced by self-steepening effect but opposite in sign. Regarding
the Raman effect, there is reflection symmetry (asymmetry) to the
gain at zero perturbation frequency when the values of the Raman coefficients
in each directional coupler are equal and with same (opposite) sign.
For the anomalous group velocity dispersion regime we observe that
the growth of the saturation parameter is followed by an increase
in the null gain region near $\Omega=0$, enlarging the separation
of the instability bands close to this point. Finally, we show that
an efficient control of the modulation instability can be realized
by adjusting self-steepening effect and intrapulse Raman scattering,
even in the presence of a saturable nonlinearity.
\end{abstract}


\maketitle

\section{Introduction}

Modulational instability (MI) is a ubiquitous phenomenon associated
with an exponential growth of the amplitude of a weak perturbed continuous
wave in certain conditions during its propagation under the interplay
between diffraction (in spatial domain) or dispersion (in temporal
domain) and nonlinearity. This phenomenon has been observed and/or
predicted in various nonlinear systems such as plasma waves \cite{Hasegawa_PRA70,McKinstrie_PFB92,Sprangle_PRL94,Gu=0000E9rin_PP95},
hydromagnetic waves \cite{Mjolhus_JPP76}, optical fibers \cite{Tai_PRL86,Agrawal_PRL87},
dust-acoustic and dust-ion-acoustic waves \cite{Amin_PRE98}, noninstantaneous
nonlinear media \cite{Kip_SCI00,Soljacic_PRL00}, Bose-Einstein condensates
\cite{Konotop_PRA02,Salasnich_PRL03,Carr_PRL04,Li_PRA05}, liquid
crystals \cite{Peccianti_NAT04}, ocean waves \cite{Onorato_PRL06,Onorato_PRL09},
and so on. In nonlinear optics, MI has been studied in lossy fibers
\cite{Karlsson_JOSAB95}, fiber gratings \cite{Russel_JP394}, for
incoherent light \cite{Kip_SCI00}, and second harmonic generation
\cite{Trillo_OL95}; and with different type of the nonlinear response
such as integrating \cite{Streppel_PRL05}, nonlocal \cite{Krolikowski_PRE01},
quadratic \cite{Schiek_PRL01}, cubic-quintic \cite{Wong_OC02}, varying
\cite{Abdullaev_JOSAB97}, and saturable one \cite{Hickmann_OL93,Dalt_OC95},
and so on. In addition, a train of solitons can emerge from a system
that presents MI allowing us to generate subpicosecond solitonlike
optical pulses, such as those presented in \cite{Hasegawa_OL84,Tai_APL86}.
A historical review on this subject can be found in \cite{Zakharova_PD09}.

Coupled systems is also a good candidate for the investigation of
MI and several studies have been proposed in this type of systems,
such as the parametric amplification and MI in dispersive nonlinear
directional couplers made from two (or more) coupled guiding channels
(e.g., two physically distinct waveguides or two polarization modes)
exhibiting an intensity-dependent refractive index with relaxing nonlinearity
\cite{Trillo_JOSAB89}, wide beam stabilities and instabilities in
one dimensional arrays of Kerr-nonlinear channel waveguides \cite{Meier_OER05},
MI in two-core optical fibers incorporating the effects of coupling-coefficient
dispersion \cite{Li_JOSAB11}, MI in birefringent two-core optical
fibers \cite{Li_JPB12}, the role of the coupling-induced group velocity
dispersion on the MI in a silicon-on-insulator directional coupler
\cite{Ding_OL12}, MI in a twin-core fiber with the effect of saturable
nonlinear response and coupling coefficient dispersion \cite{Nithyanandan_PRA13},
the investigation of the dynamic properties of a nonlinear directional
coupler made of Kerr materials inducing MI was presented in Ref. \cite{Ogusu_JLT13},
the interplay between relaxation of nonlinear response and coupling
coefficient dispersion in the MI of dual core optical fiber \cite{Nithyanandan_OC13},
MI in an array of positive- and negative-index waveguides \cite{Zhang_JOSAB14},
MI in nonlinear positive-negative index couplers with saturable nonlinearity
\cite{Tatsing_JOSAB12}, MI of copropagating light beams induced by
cubic-quintic nonlinearity in nonlinear negative-index material \cite{Gupta_JOSAB12},
MI in nonlinear oppositely directed coupler with a negative-index
metamaterial channel \cite{Xiang_PRE10}, and so on. Furthermore,
some systems can exhibit significant changes in the spectrum of the
MI due to higher order effects \cite{Mohamadou_JMO14,Ali_PRE14,Reyna_PRA14,Nithyanandan_PJP14,Zhong_O14,Hu_LPL13,Nithyanandan_PRA12,Ganapathy_PJP02}.
To be specific, the influence of self-steepening and intrapulse Raman
scattering on MI in oppositely directed coupler was studied \cite{Ali_PRE14}.

Regarding to the effects of a nonlinear saturation present in some
physical systems of interest, the band of instability may present
significant changes in its amplitude and/or shape, and may even vanishing.
This has motivated to investigate the MI in several saturable nonlinear
systems such as waveguides \cite{Dalt_OC95,Stepic_OC06}, optical
fibers \cite{Lyra_OC94,Nithyanandan_PRA13}, negative refractive metamaterial
\cite{Zhong_OC11}, metamaterials \cite{Zhong_O14,Xiang_JOSAB11,Zhong_JOSAB14},
optical fibers with higher-order dispersion \cite{Nithyanandan_PJP14,Dinda_JOSAB10,Porsezian_JOSAB12},
fibers with saturable delayed nonlinear response \cite{Silva_JOSAB09},
positive-negative index couplers \cite{Tatsing_JOSAB12}, semiconductor-doped
glass fibers \cite{Hickmann_OL93}, and liquid-core photonic crystal
fibers \cite{Raja_PRA10}.

The aim of the present work is to investigate the influence of a saturable
nonlinearity on the MI in oppositely directed coupler in the presence
of high-order effects. As a particular case of our model, in the absence
of saturation the system should present similar results to those obtained
in \cite{Ali_PRE14} for the effects of self-steepening and intrapulse
Raman scattering on the MI. 

The paper is organized as follows: We introduce the theoretical model
and present the analytical results for the power gain in the next
section; in Sec. \ref{sec:Numerical-results} we display the numerical
results, in which we check the influence of self-steepening and intrapulse
Raman scattering on the MI in Subsecs. \ref{sub:Effect-of-self-steepening}
and \ref{sub:Effect-of-intrapulse}, respectively. Our conclusions
are shown in Sec. \ref{sec:Conclusion}.

\section{Theoretical model}

The model that describes the propagation of a high intense optical
beam in oppositely directed coupler is given by the pair of linearly
coupled nonlinear Schr\"odinger equations (CNLSE) with the form \cite{Ali_PRE14,Nithyanandan_PRA13}:\begin{subequations}
\begin{multline}
i\sigma_{1}\frac{\partial u_{1}}{\partial z}-\frac{\beta_{21}}{2}\frac{\partial^{2}u_{1}}{\partial x^{2}}+\kappa_{12}u_{2}e^{-i\delta z}+\gamma_{1}\{f(\Gamma|u_{1}|^{2})u_{1}+\\
is_{1}\frac{\partial(f(\Gamma|u_{1}|^{2})u_{1})}{\partial x}-T_{R1}u_{1}\frac{\partial(f(\Gamma|u_{1}|^{2}))}{\partial x}\}=0,\label{u1}
\end{multline}
\begin{multline}
i\sigma_{2}\frac{\partial u_{2}}{\partial z}-\frac{\beta_{22}}{2}\frac{\partial^{2}u_{2}}{\partial x^{2}}+\kappa_{21}u_{1}e^{i\delta z}+\gamma_{2}\{f(|\Gamma u_{2}|^{2})u_{2}\\
+is_{2}\frac{\partial(f(\Gamma|u_{2}|^{2})u_{2})}{\partial x}-T_{R2}u_{2}\frac{\partial(f(|\Gamma u_{2}|^{2}))}{\partial x}\}=0,\label{u2}
\end{multline}
\end{subequations}where $\sigma_{1}$ and $\sigma_{2}$ indicate
the sign of refractive index in channel-$1$ and channel-$2$ of the
coupler, respectively. In order to compare the results obtained by
our model with those presented in Ref. \cite{Ali_PRE14}, we consider
here the channel-$1$ made by positive index material and channel-$2$
by negative index material, hence $\sigma_{1}=1$ and $\sigma_{2}=-1$;
$\beta_{21}$ and $\beta_{22}$ are group velocity dispersion coefficients,
$u_{1}(z,x)$ and $u_{2}(z,x)$ stand for the normalized complex amplitude
of the modes in channels \emph{$1$} and $2$, $\kappa_{12}$ and
$\kappa_{21}$ are linear coupling coefficients, $\delta=\beta_{1}-\beta_{2}$,
where $\beta_{1}$ and $\beta_{2}$ represent the propagation constants
of the individual channels; $\gamma_{1}$ and $\gamma_{2}$ are the
nonlinear coefficients related to self-phase modulation, $s_{1}$
and $s_{2}$ represent self-steepening effects, and $T_{R1}$ and
$T_{R2}$ are responsible for the Raman-induced frequency shift, induced
by intrapulse Raman scattering. Also, in this model we neglect the
cross-phase modulation effects.

To be more specific, in the next steps, the functions $f(\Gamma|u_{1}|^{2})$
and $f(\Gamma|u_{2}|^{2})$ standing for the dependence of the refractive
index with the intensity of the incident radiation will be described
by a model widely used in the literature describing the saturable
nonlinearity as follows \cite{Dinda_JOSAB10,Chen_JOSAB06,Gatz_JOSAB91}
\begin{equation}
f(\Gamma|u_{j}|^{2})=\frac{|u_{j}|^{2}}{1+\Gamma|u_{j}|^{2}},
\end{equation}
where $\Gamma=1/P_{S}$ is the saturation parameter with $P_{s}$
being the saturation power.

\subsection{Modulation instability}

\begin{figure*}
\centering\includegraphics[width=0.48\columnwidth]{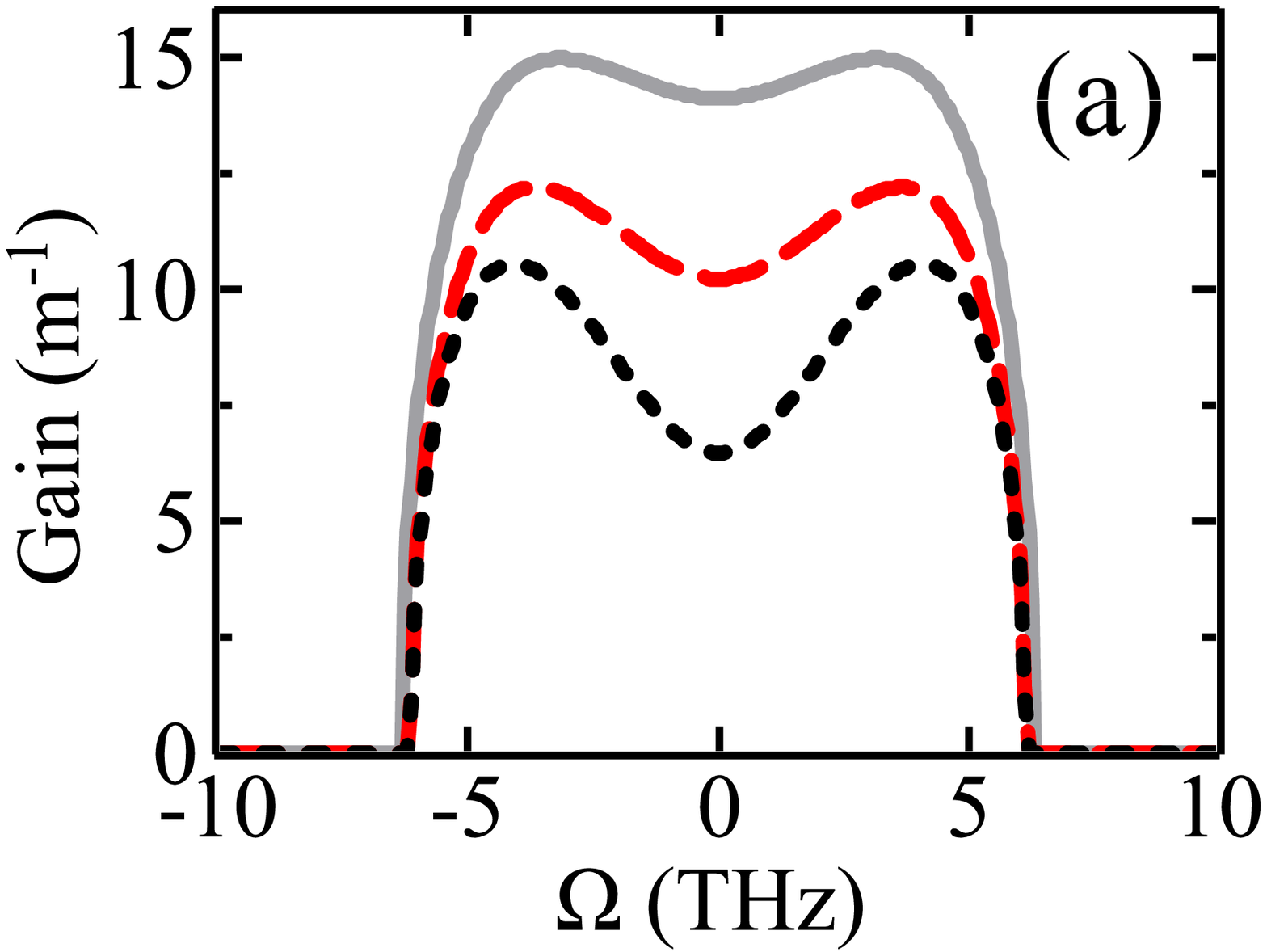} \includegraphics[width=0.48\columnwidth]{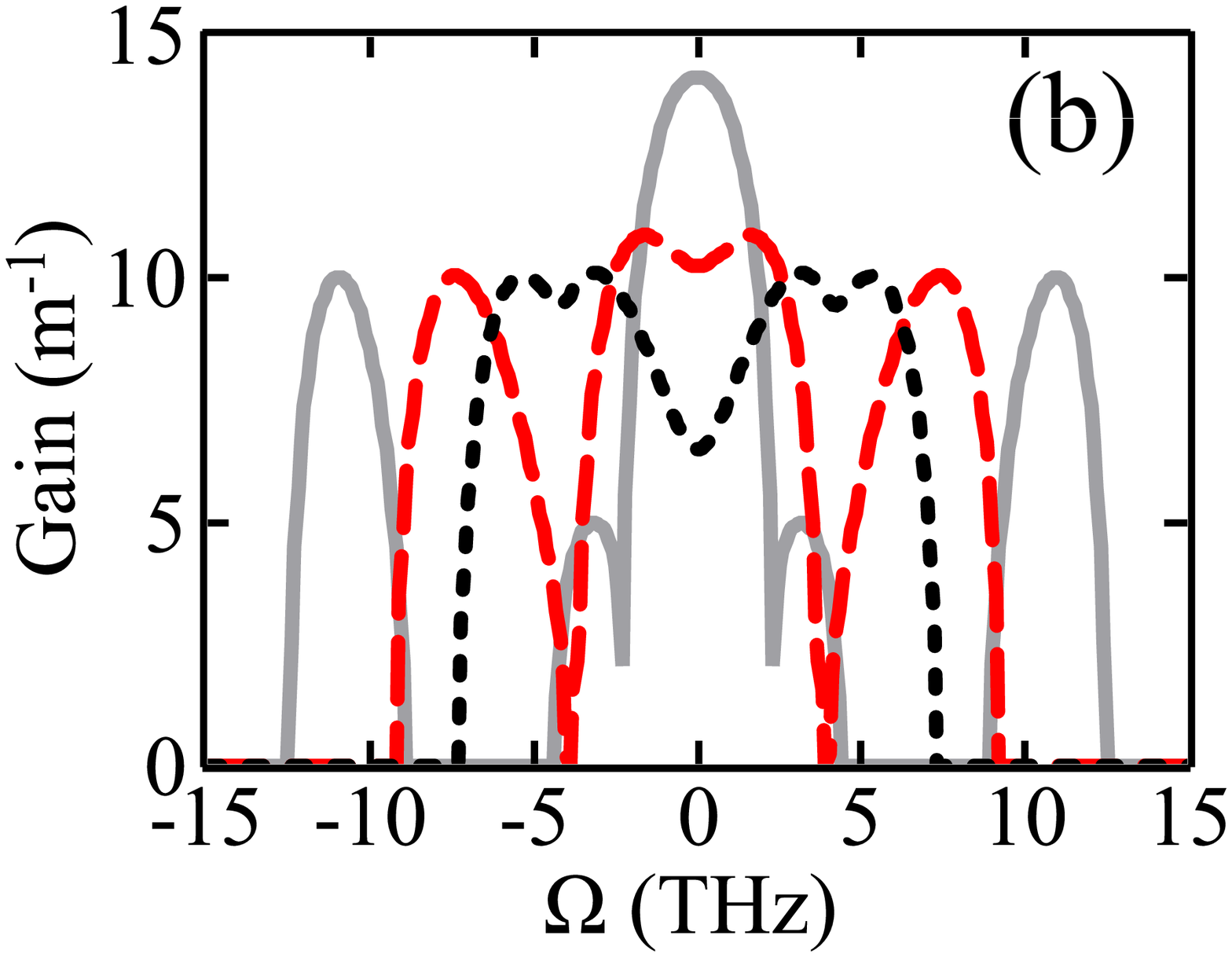}
\includegraphics[width=0.48\columnwidth]{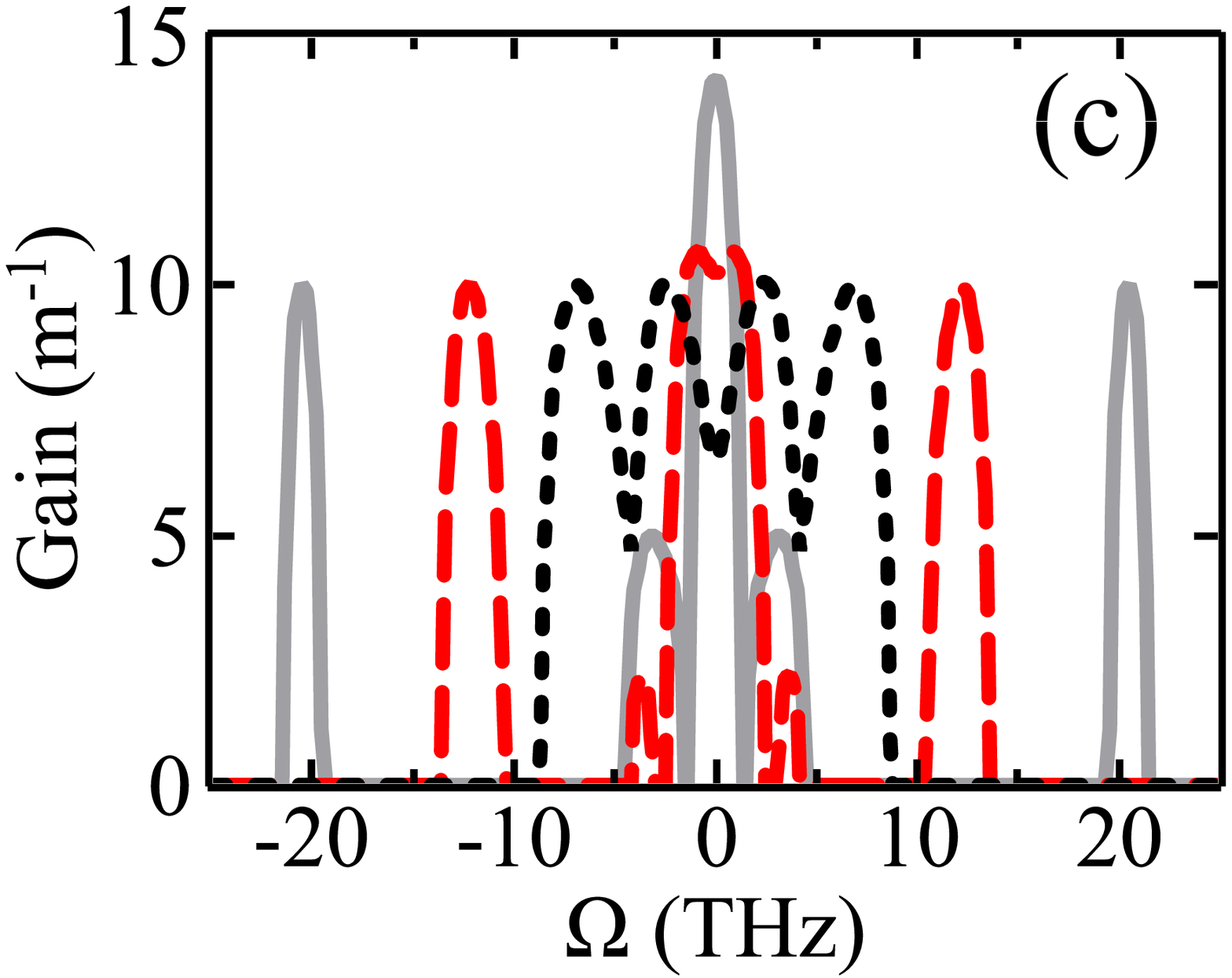} \includegraphics[width=0.48\columnwidth]{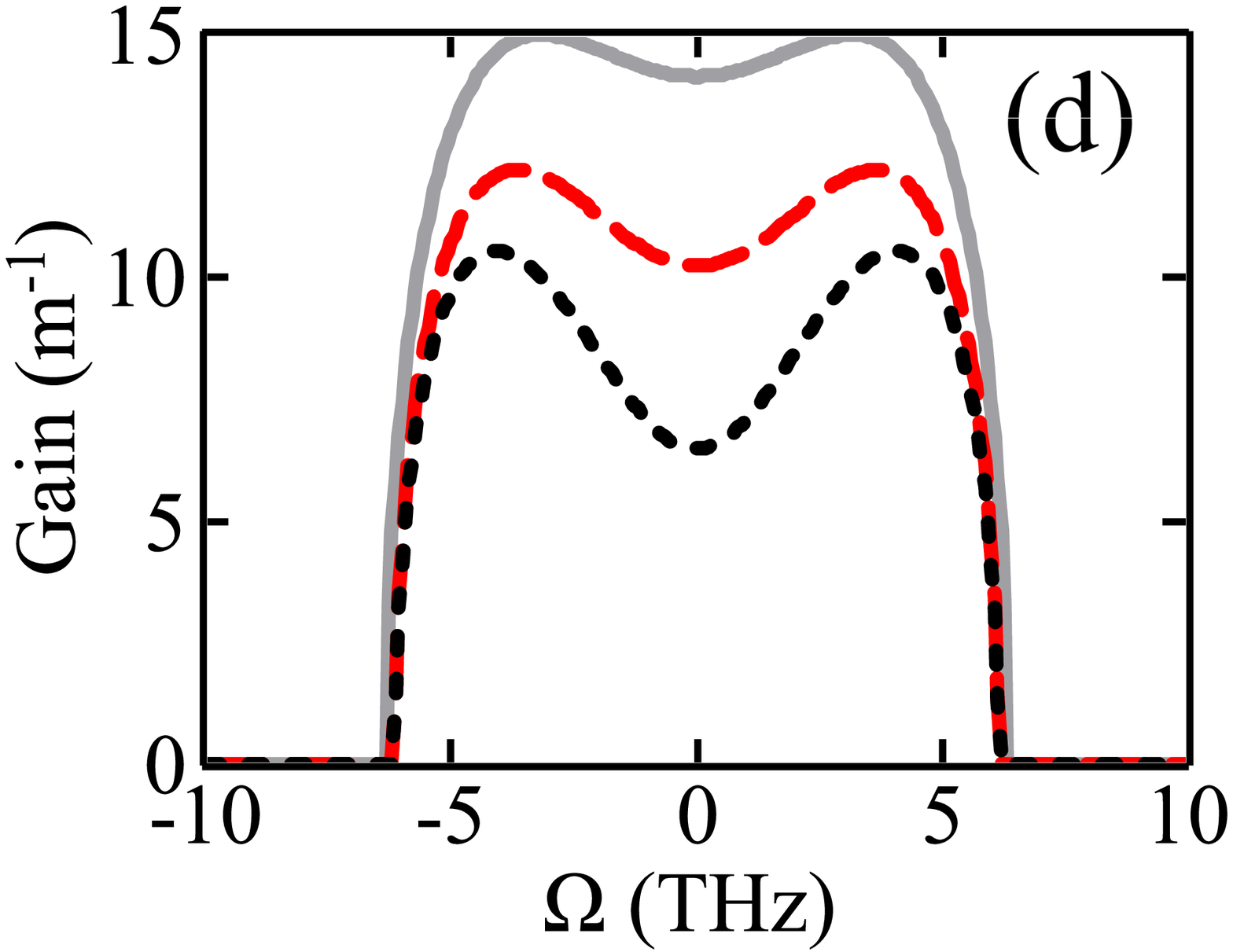}

\protect\caption{(Color online) Instability gain spectra in normal group velocity dispersion
regime as function of saturation parameter under different combinations
of $s_{1}$ and $s_{2}$ when $P=10\mbox{{kW}}$, $\gamma_{1}=\gamma_{2}=1/(\mathrm{kW\,m})$,
and $\kappa_{12}=\kappa_{21}=10\mbox{{m}}^{-1}$ with (a) $s_{1}=s_{2}=0$,
(b) $s_{1}=0$ and $s_{2}=1\:\mathrm{ps/(kW\,m)}$, (c) $s_{1}=s_{2}=1\:\mathrm{ps/(kW\,m)}$,
and (d) $s_{1}=-s_{2}=1\:\mathrm{ps/(kW\,m)}$. The saturation parameter
values used herein are: $\Gamma=0$ in solid-line (gray), $\Gamma=0.1\:\mathrm{kW^{-1}}$
in dashed-line (red), and $\Gamma=0.4\:\mathrm{kW^{-1}}$ in dotted-line
(black). The other parameters are $T_{R1}=T_{R2}=0\:\mathrm{ps/(kW\,m)}$,
$h=1$ and $\beta_{21}=\beta_{22}=1\:\mathrm{ps^{2}\,m^{-1}}$.}

\label{F1}
\end{figure*}

\begin{figure*}[tb]
\centering\includegraphics[width=0.48\columnwidth]{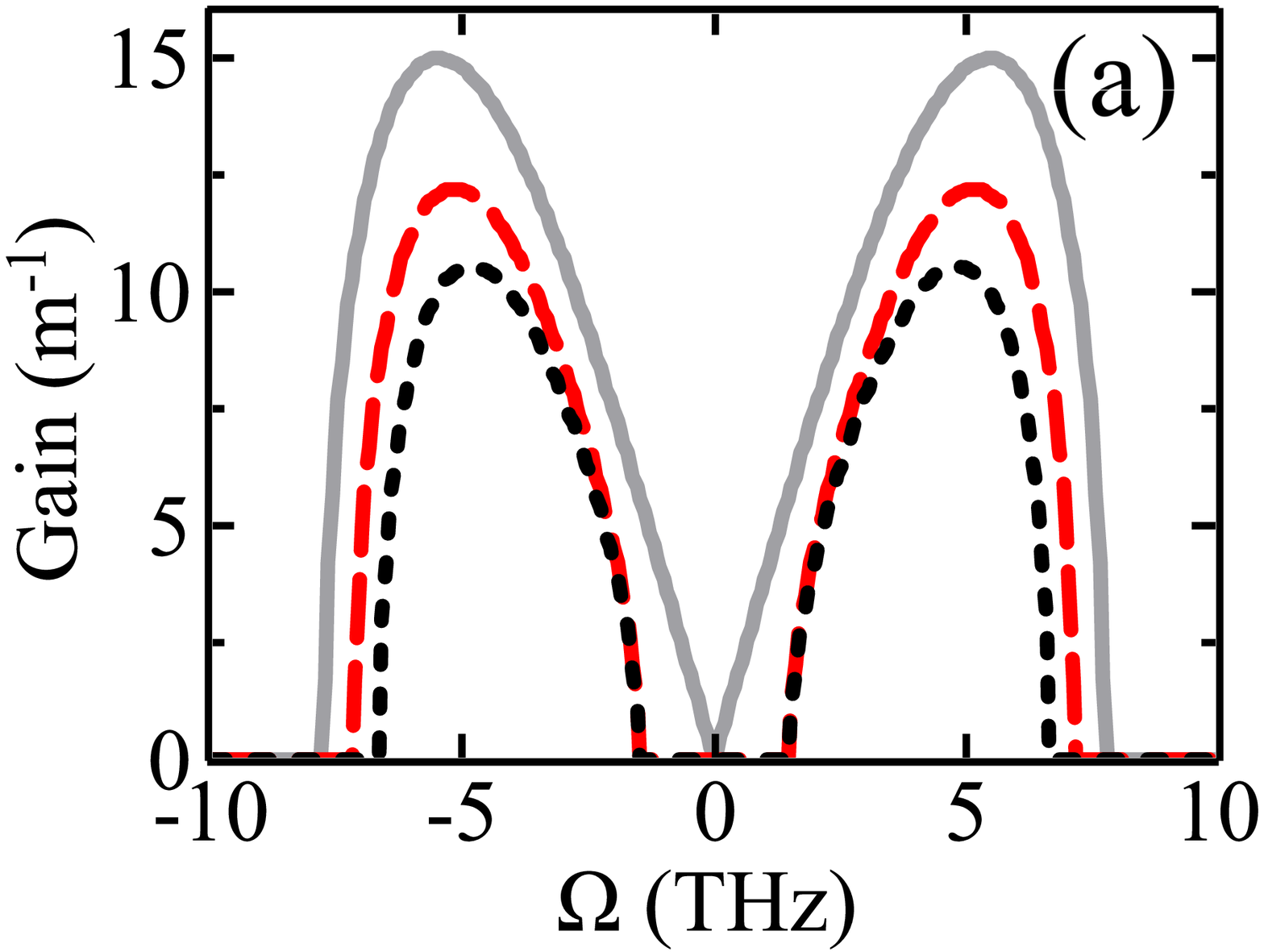} \includegraphics[width=0.48\columnwidth]{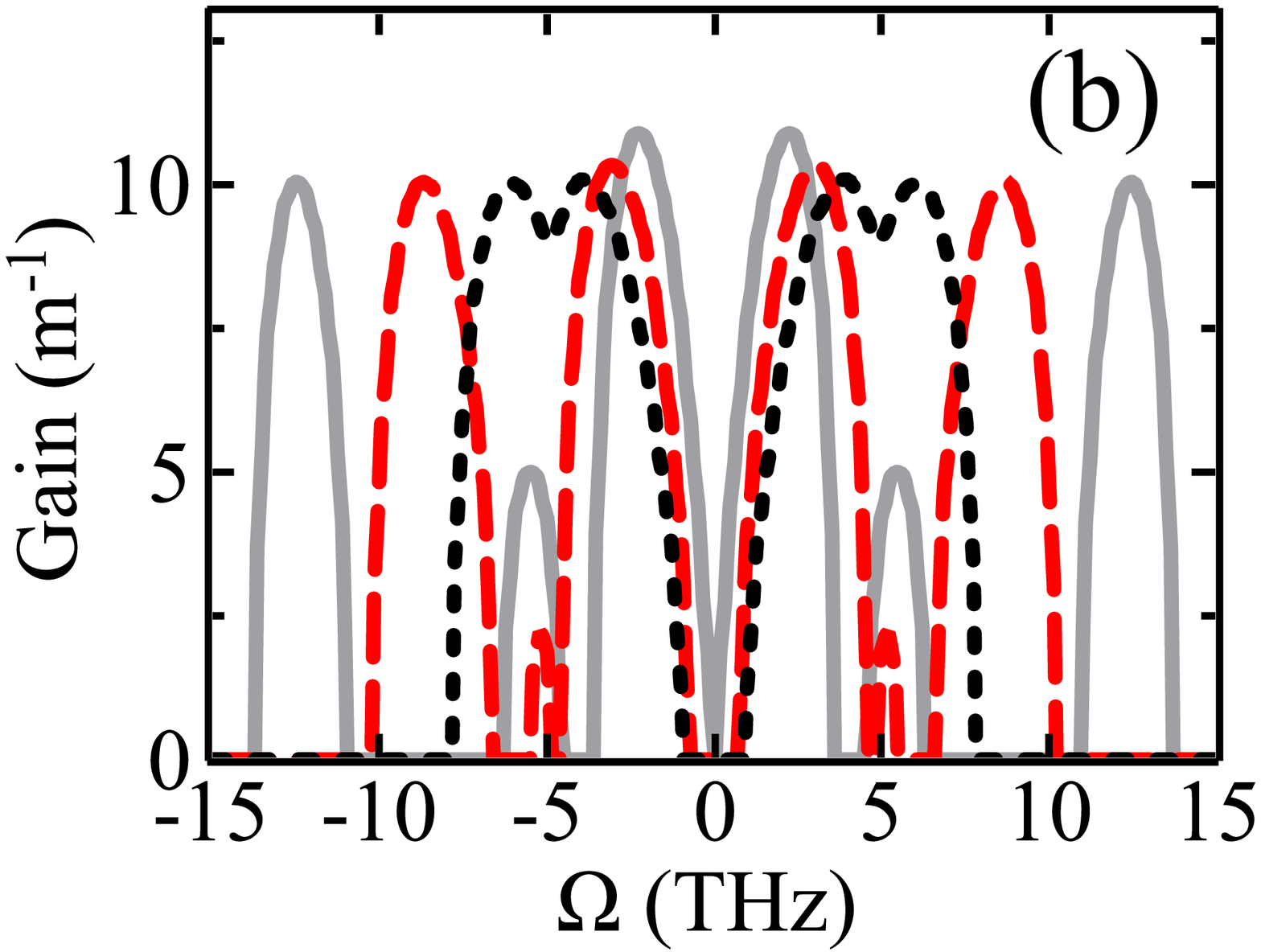}
\includegraphics[width=0.48\columnwidth]{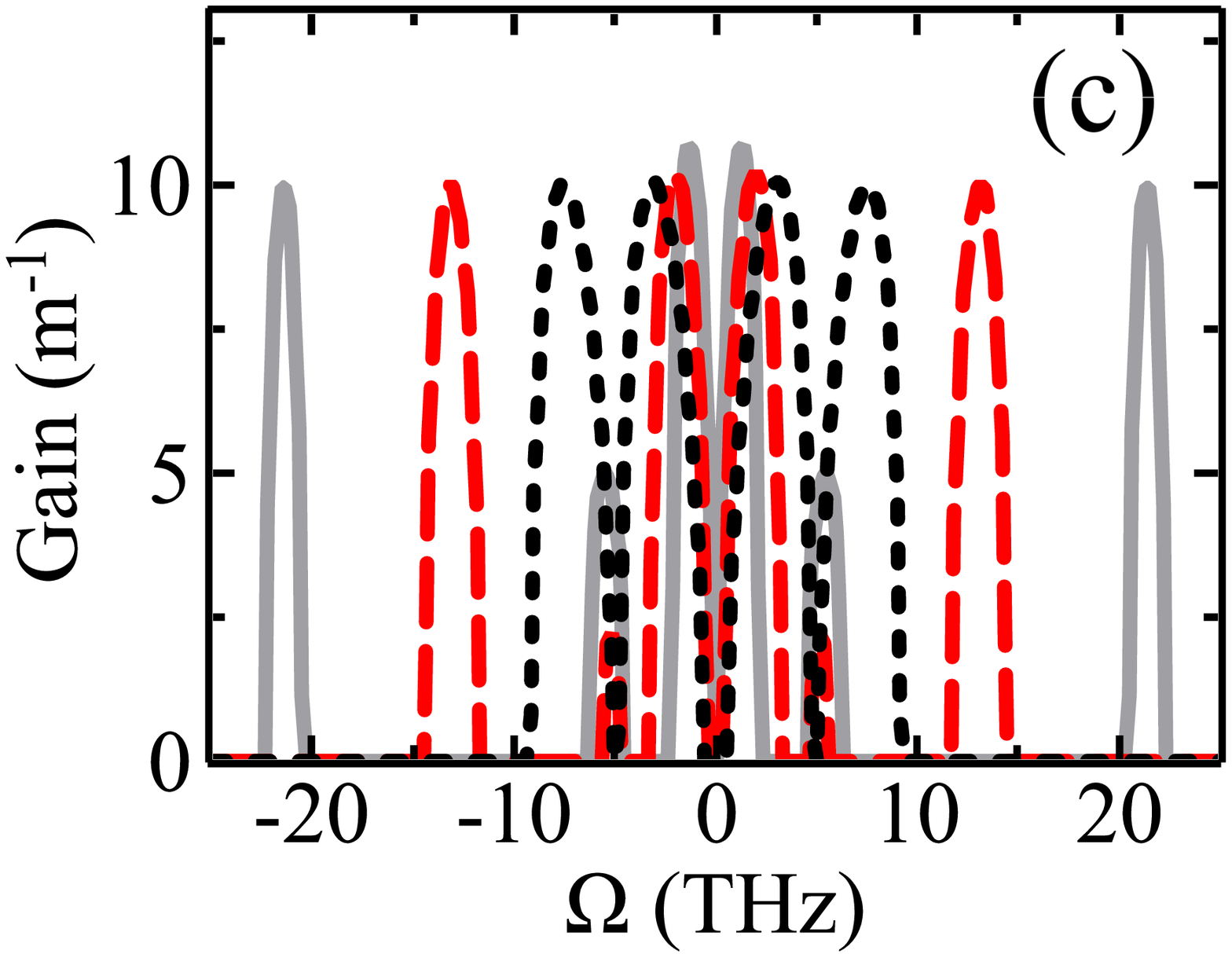} \includegraphics[width=0.48\columnwidth]{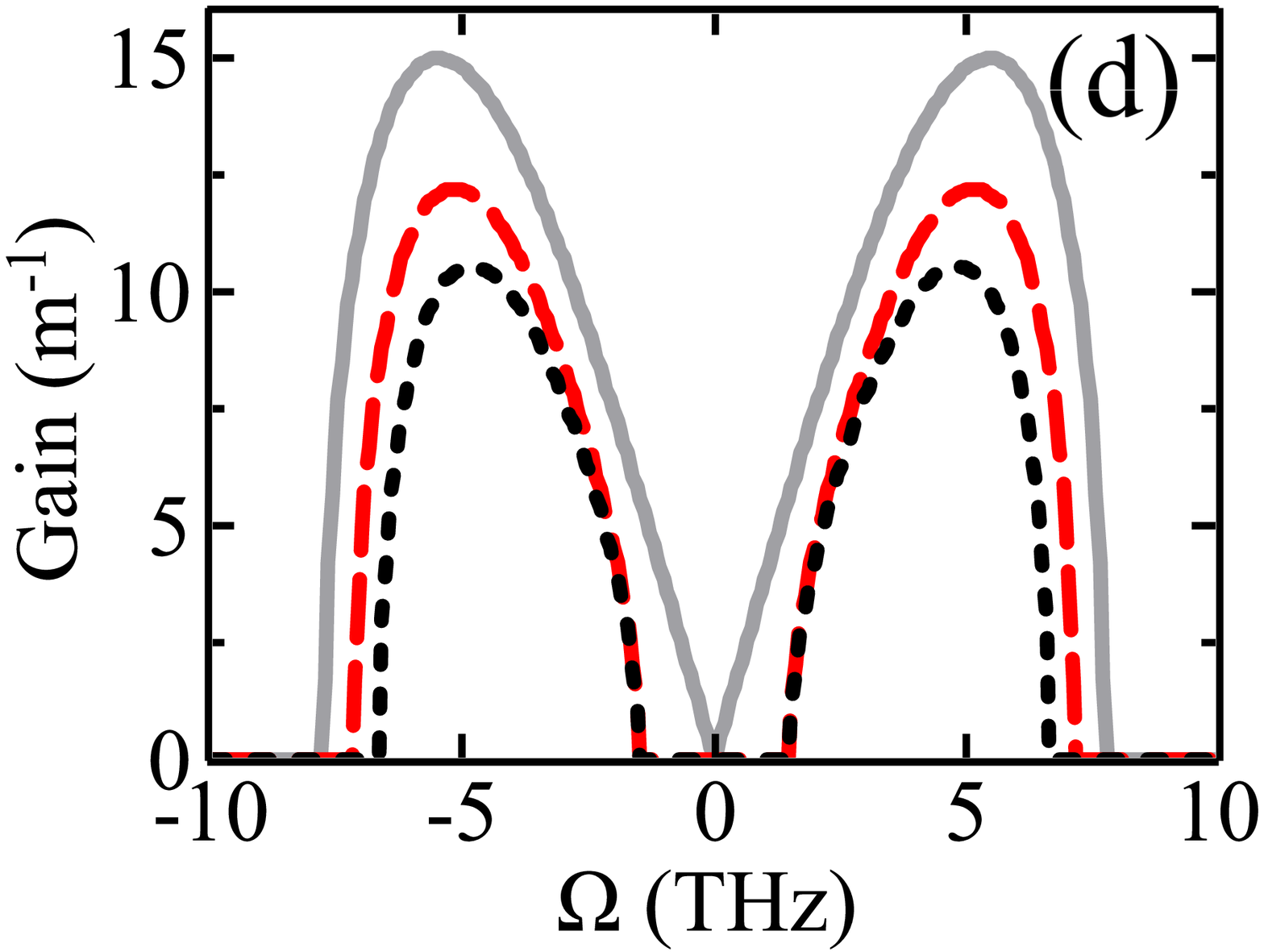}

\protect\caption{(Color online) Instability gain spectra in anomalous group velocity
dispersion regime as function of saturation parameter under different
combinations of $s_{1}$ and $s_{2}$ when $P=10\mbox{{kW}}$, $\gamma_{1}=\gamma_{2}=1/(\mathrm{kW\,m})$,
and $\kappa_{12}=\kappa_{21}=10\mbox{{m}}^{-1}$ with (a) $s_{1}=s_{2}=0$,
(b) $s_{1}=0$ and $s_{2}=1\:\mathrm{ps/(kW\,m)}$, (c) $s_{1}=s_{2}=1\:\mathrm{ps/(kW\,m)}$,
and (d) $s_{1}=-s_{2}=1\:\mathrm{ps/(kW\,m)}$. The saturation parameter
values used herein are: $\Gamma=0$ in solid-line (gray), $\Gamma=0.1\:\mathrm{kW^{-1}}$
in dashed-line (red), and $\Gamma=0.4\:\mathrm{kW^{-1}}$ in dotted-line
(black). The other parameters are $T_{R1}=T_{R2}=0\:\mathrm{ps/(kW\,m)}$,
$h=-1$ and $\beta_{21}=\beta_{22}=-1\:\mathrm{ps^{2}\,m^{-1}}$.}

\label{F2}
\end{figure*}

\begin{figure*}[tb]
\includegraphics[width=0.65\columnwidth]{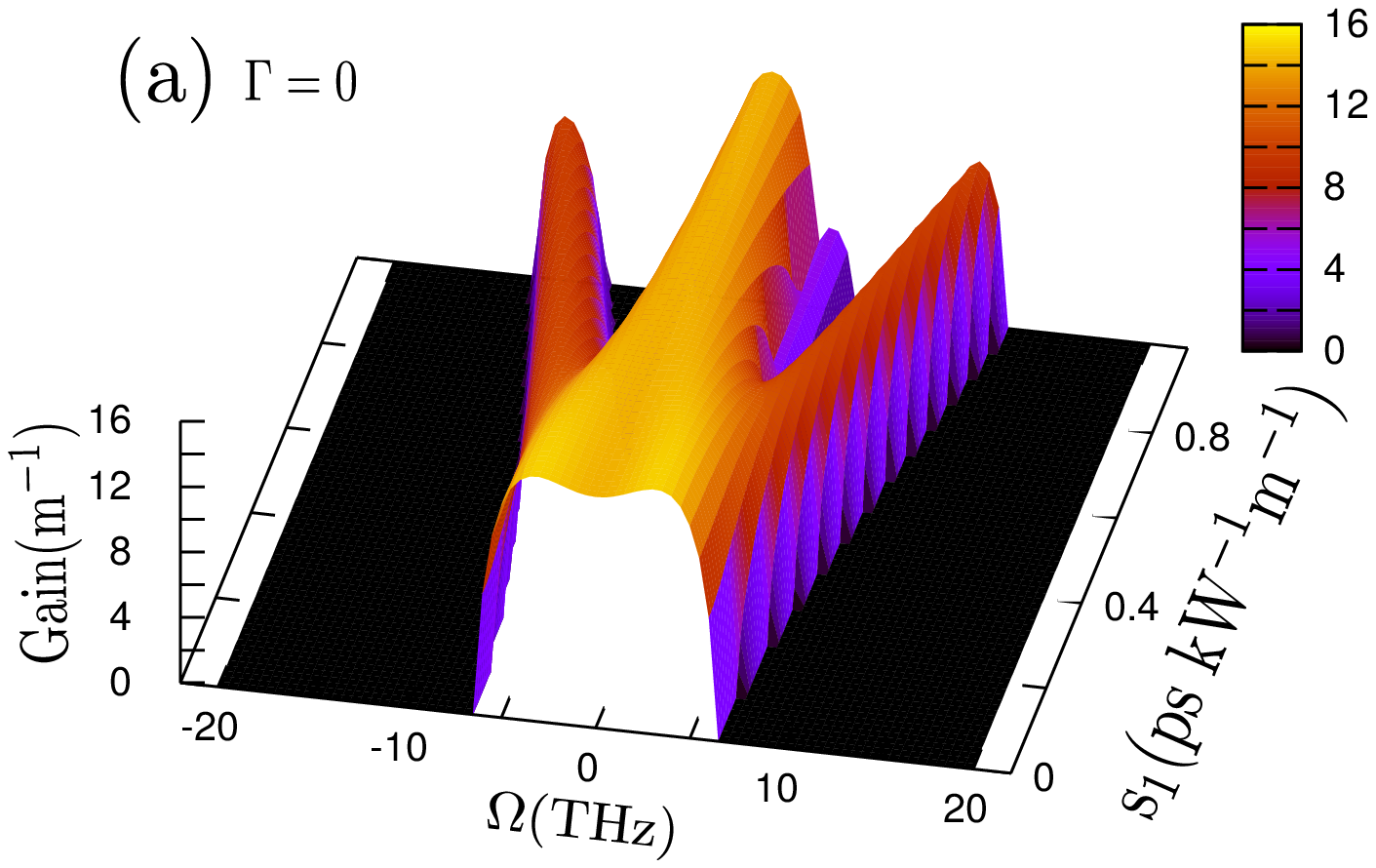} \includegraphics[width=0.65\columnwidth]{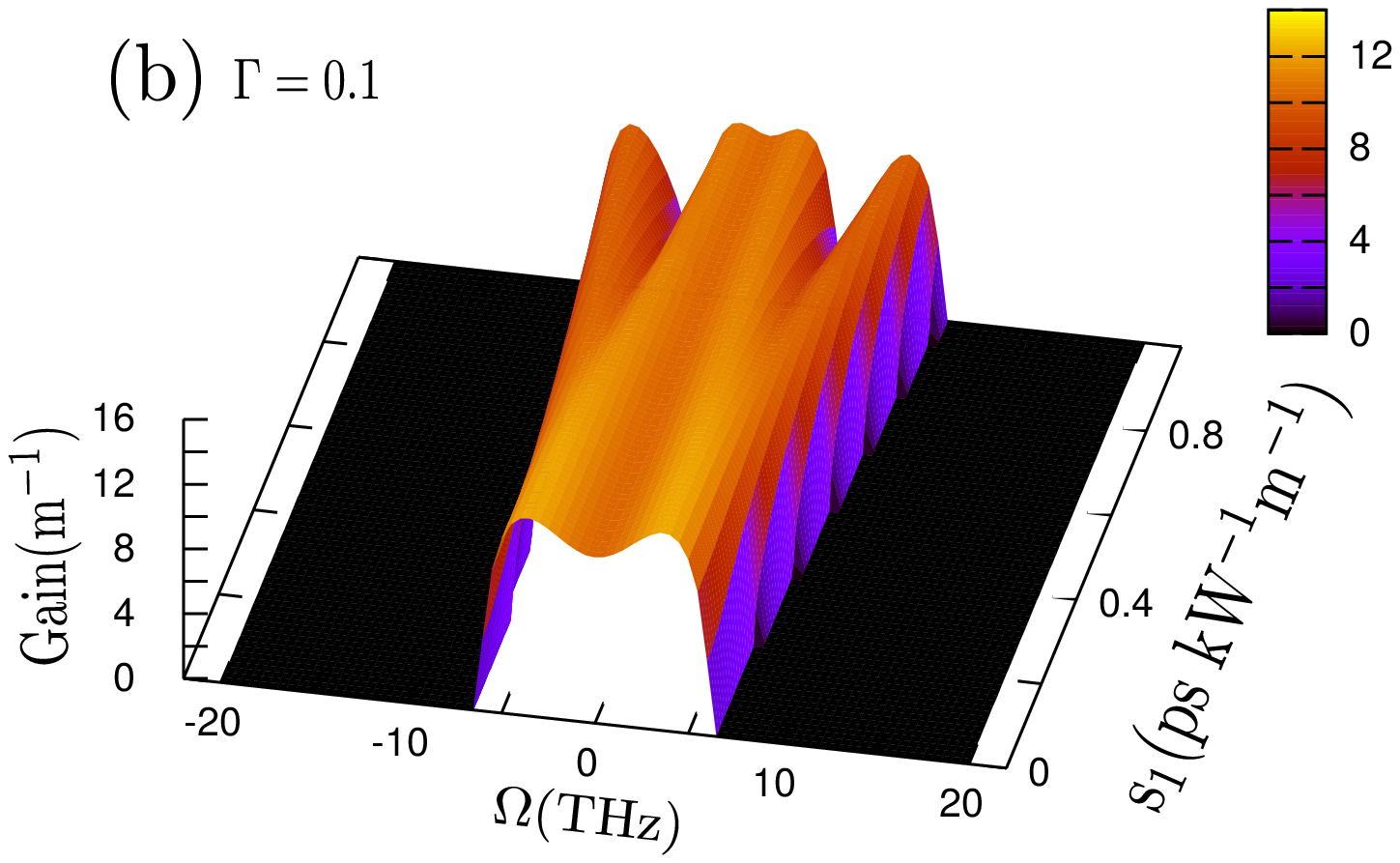}
\includegraphics[width=0.65\columnwidth]{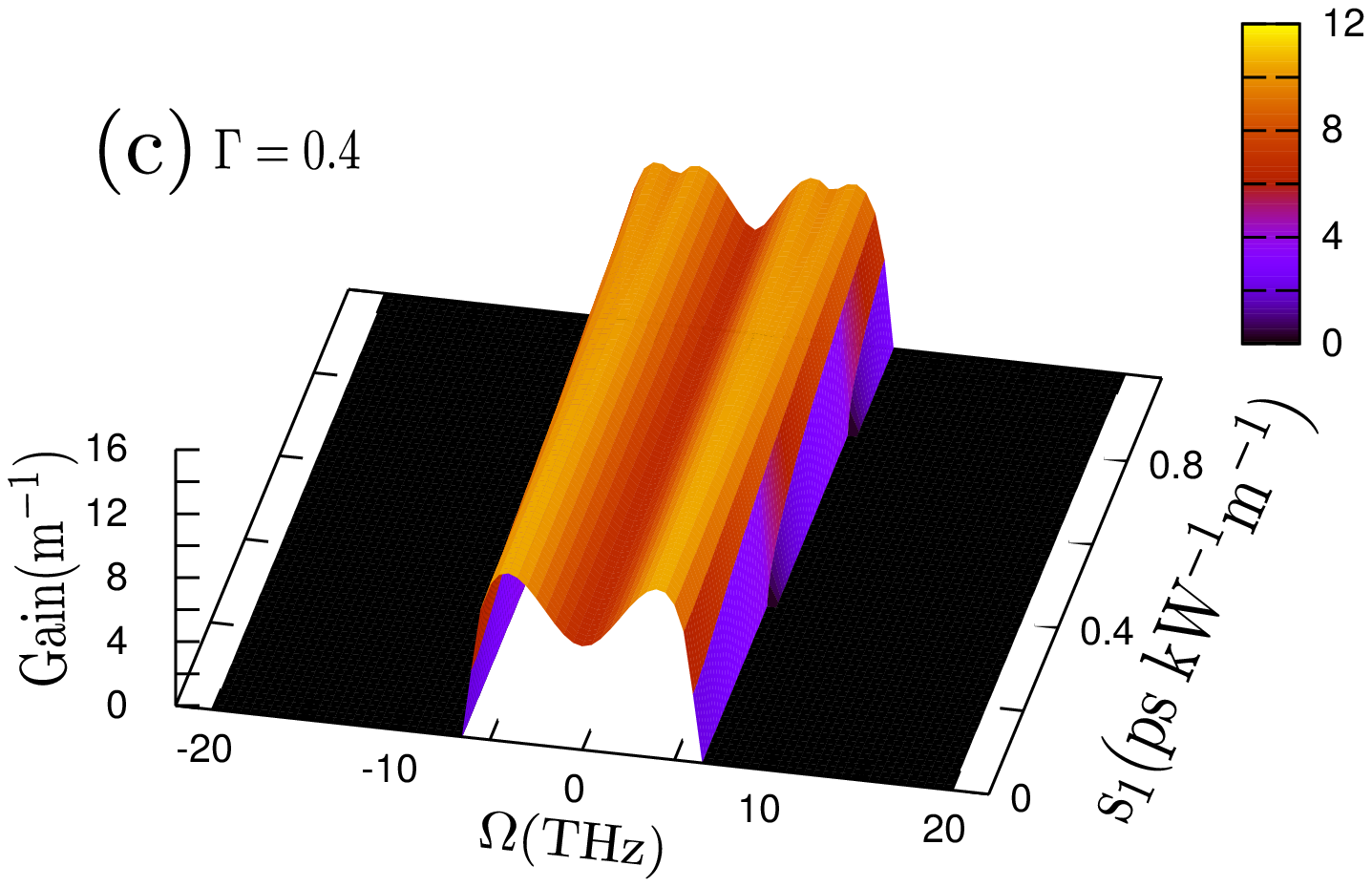}

\protect\caption{(Color online) Instability gain spectra showing the influence of the
self-steepening effect in channel $1$ ($s_{2}=0$) in normal group
velocity dispersion regime for different values of saturation parameter.
The other parameters are the same ones used in Fig. \ref{F1}.}

\label{F3}
\end{figure*}

\begin{figure*}[tb]
\includegraphics[width=0.65\columnwidth]{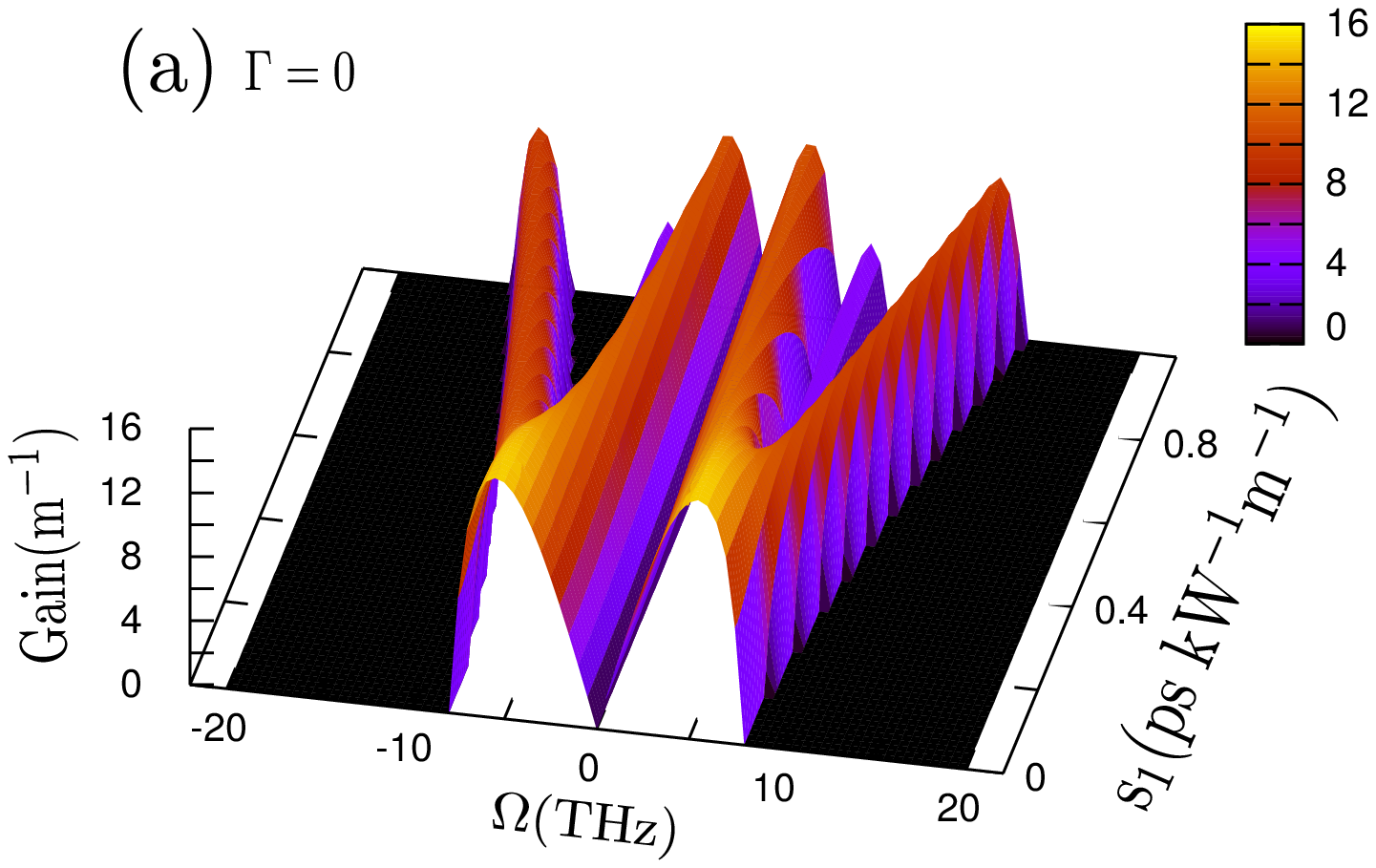} \includegraphics[width=0.65\columnwidth]{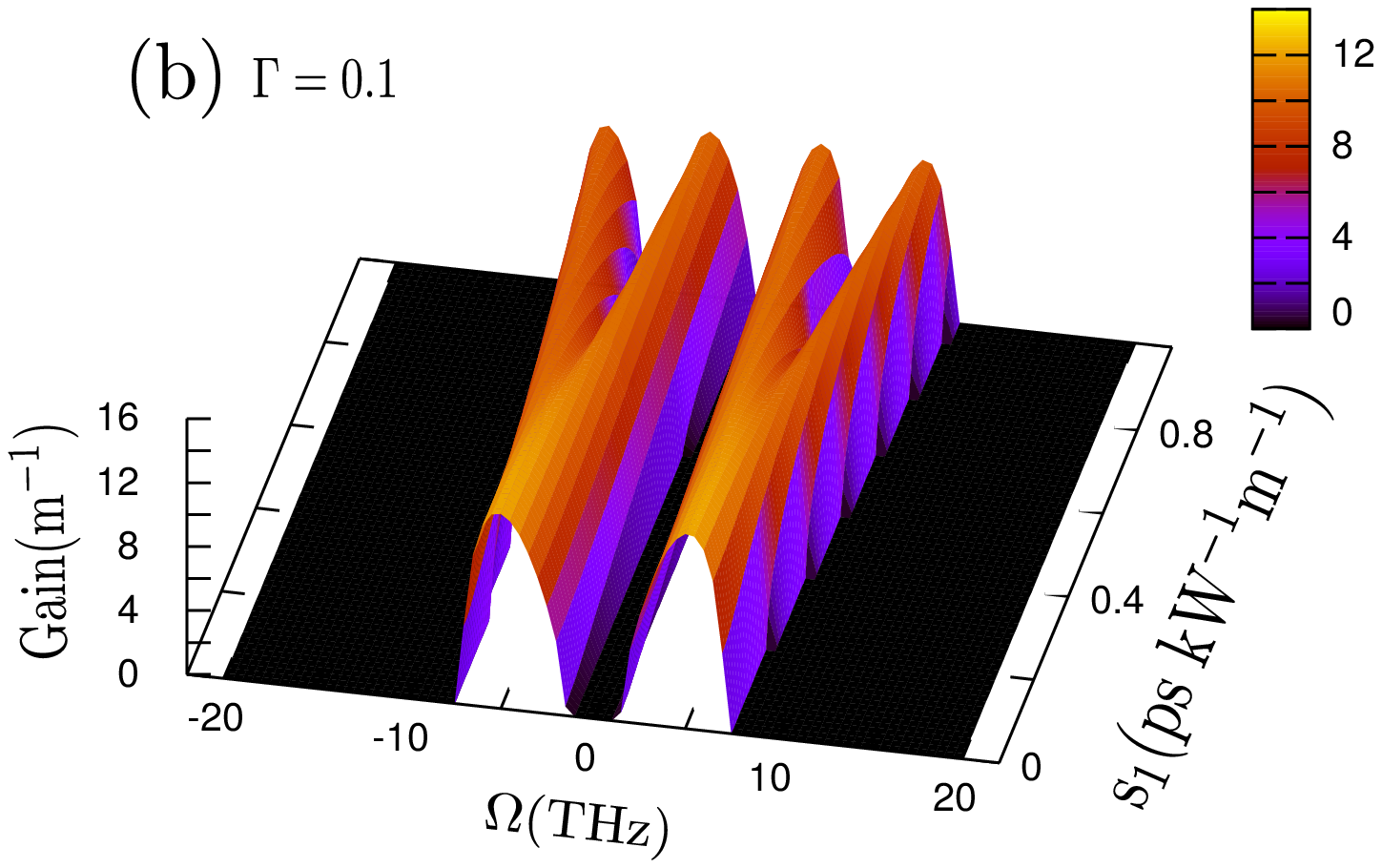}
\includegraphics[width=0.65\columnwidth]{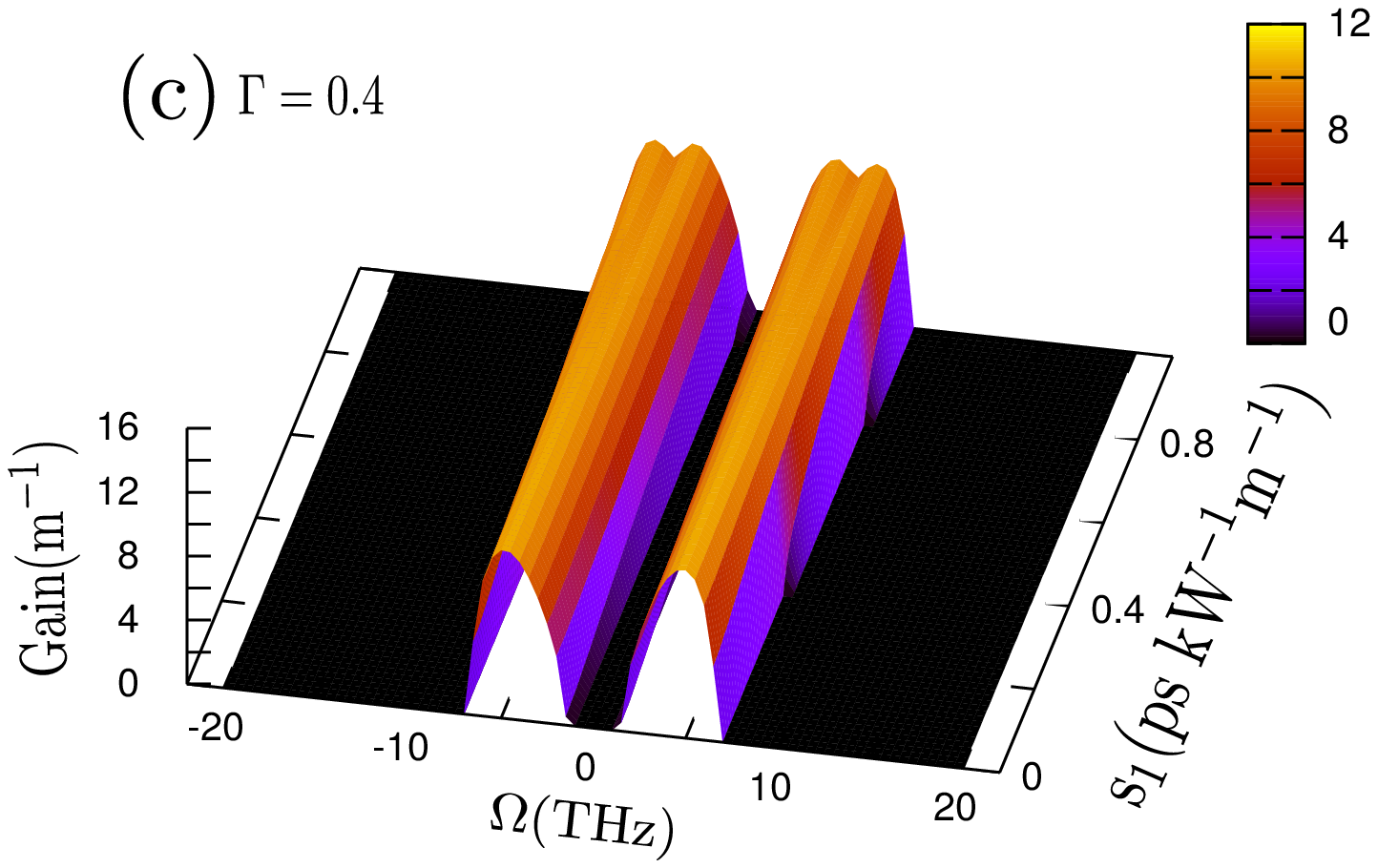}

\protect\caption{(Color online) Instability gain spectra showing the influence of the
self-steepening effect in channel $1$ ($s_{2}=0$) in anomalous group
velocity dispersion regime for different values of saturation parameter.
The other parameters are the same ones used in Fig. \ref{F2}.}

\label{F4}
\end{figure*}

In order to verify the influence of a saturable nonlinearity on the
MI we use the standard linear stability analysis. The basic idea of
linear stability analysis is to perturb a continuous wave solution
and then analyze whether this small perturbation grows or decays with
propagation. For this, consider that the steady-state solutions of
Eqs. (\ref{u1}) and (\ref{u2}) can be written as\begin{subequations}
\begin{equation}
u_{1}(z)=a_{1}e^{iqz}e^{-i\frac{\delta}{2}z},\label{u1cw}
\end{equation}
\begin{equation}
u_{2}(z)=a_{2}e^{iqz}e^{i\frac{\delta}{2}z},\label{u2cw}
\end{equation}
\end{subequations}with $a_{j}=\sqrt{P_{j}}$ , and $P_{j}$ is the
input power in the coupler $j=1,2$. By inserting the Eqs. (\ref{u1cw})
and (\ref{u2cw}) in Eqs. (\ref{u1}) and (\ref{u2}) one obtains
the terms $q$ and $\delta$ given by\begin{subequations}
\begin{equation}
q=\frac{1}{2}\left[\kappa_{12}h-\kappa_{21}h^{-1}+\frac{\gamma_{1}R}{1+\Gamma R}-\frac{\gamma_{2}Rh^{2}}{1+\Gamma Rh^{2}}\right],
\end{equation}
\begin{equation}
\delta=-\left[\kappa_{21}h^{-1}+\kappa_{12}h+\frac{\gamma_{1}R}{1+\Gamma R}+\frac{\gamma_{2}Rh^{2}}{1+\Gamma Rh^{2}}\right],
\end{equation}
\end{subequations}where $h\equiv a_{2}/a_{1}$, which describes how
the total power $P=a_{1}^{2}+a_{2}^{2}$ is divided between forward
and backward propagating waves, and $R\equiv P/(1+h^{2})$.

Next, we assume the stead-state solutions (\ref{u1cw}) and (\ref{u2cw})
can be perturbed by the functions $\alpha_{j}(z,x)$, such that\begin{subequations}
\begin{equation}
u_{1}(z,x)=\left[a_{1}+\alpha_{1}(z,x)\right]e^{iqz}e^{-i\frac{\delta}{2}z},\label{u1pert}
\end{equation}
\begin{equation}
u_{2}(z,x)=\left[a_{2}+\alpha_{2}(z,x)\right]e^{iqz}e^{i\frac{\delta}{2}z},\label{u2pert}
\end{equation}
\end{subequations}where $\alpha_{i}(z,x)$ is a small perturbation
satisfying $|\alpha_{i}(z,x)|\ll\sqrt{P_{i}}$. Now, consider the
perturbation terms as combinations of plane waves with the following
form
\begin{equation}
\alpha_{j}(z,x)=c_{j}e^{i[Kz-\Omega x]}+d_{j}e^{-i[Kz-\Omega x]},\label{pert}
\end{equation}
where $K$ and $\Omega$ are wave-vector and frequency of perturbation
amplitude. Thus, inserting the Eqs.(\ref{u1pert}) and (\ref{u2pert})
in (\ref{u1}) and (\ref{u2}), after some mathematical manipulations,
one obtains the linearized equations \begin{subequations}
\begin{multline}
i\frac{\partial\alpha_{1}}{\partial z}-\frac{\beta_{21}}{2}\frac{\partial^{2}\alpha_{1}}{\partial x^{2}}+\frac{\gamma_{1}R}{(1+\Gamma R)^{2}}\left[\alpha_{1}(1+\Gamma R)+\alpha_{1}^{*}\right]\\
-\alpha_{1}\kappa_{12}h+\alpha_{2}\kappa_{12}+\frac{i\gamma_{1}s_{1}R}{(1+\Gamma R)^{2}}\left[\frac{\partial\alpha_{1}}{\partial x}(2+\Gamma R)\right]\\
-\frac{\gamma_{1}T_{R1}R}{(1+\Gamma^{2}+2\Gamma R)}\frac{\partial\alpha_{1}}{\partial x}=0,\label{eqlinear01}
\end{multline}

\begin{multline}
-i\frac{\partial\alpha_{2}}{\partial z}-\frac{\beta_{22}}{2}\frac{\partial^{2}\alpha_{2}}{\partial x^{2}}+\frac{\gamma_{2}Rh^{2}}{(1+\Gamma Rh^{2})^{2}}\left[\alpha_{2}(1+\Gamma Rh^{2})+\alpha_{2}^{*}\right]\\
-\alpha_{2}\kappa_{21}h^{-1}+\alpha_{1}\kappa_{21}+\frac{i\gamma_{2}s_{2}Rh^{2}}{(1+\Gamma Rh^{2})^{2}}\left[\frac{\partial\alpha_{2}}{\partial x}(2+\Gamma Rh^{2})\right]\\
-\frac{\gamma_{2}T_{R2}Rh^{2}}{(1+\Gamma^{2}+2\Gamma Rh^{2})}\frac{\partial\alpha_{2}}{\partial x}=0.\label{eqlinear02}
\end{multline}
\end{subequations}Substituting Eq. (\ref{pert}) into Eqs. (\ref{eqlinear01})
and (\ref{eqlinear02}), one gets a set of four linearly coupled equations
satisfied by $c_{j}$ and $d_{j}$. This set of coupled equations
can be written in matrix form\[\left(\begin{array}{llll}
m_{11} & m_{12} & m_{13} & m_{14} \\
m_{21} & m_{22} & m_{23} & m_{24} \\
m_{31} & m_{32} & m_{33} & m_{34} \\
m_{41} & m_{42} & m_{43} & m_{44}
\end{array}\right)
\left(\begin{array}{l}
c_1 \\ c_2 \\ d_1 \\ d_2
\end{array}\right)=0,
\]where the matrix elements are given by: $m_{11}=0$; $m_{12}=\frac{\gamma_{2}Rh^{2}}{(1+\Gamma Rh^{2})^{2}}$;
$m_{13}=\kappa_{21}$; $m_{14}=-K+\frac{\beta_{22}}{2}\Omega^{2}-\kappa_{21}h^{-1}+\frac{\gamma_{2}Rh^{2}}{(1+\Gamma Rh^{2})}-\frac{\gamma_{2}s_{2}Rh^{2}\Omega(2+\Gamma Rh^{2})}{(1+\Gamma Rh^{2})^{2}}-\frac{i\gamma_{2}T_{R2}Rh^{2}\Omega}{(1+\Gamma^{2}+2\Gamma Rh^{2})}$;
$m_{21}=\frac{\gamma_{1}R}{(1+\Gamma R)^{2}}$; $m_{22}=0$; $m_{23}=K+\frac{\beta_{21}}{2}\Omega^{2}-\kappa_{12}h+\frac{\gamma_{1}R}{(1+\Gamma R)}-\frac{\gamma_{1}s_{1}R\Omega(2+\Gamma R)}{(1+\Gamma R)^{2}}-\frac{i\gamma_{1}T_{R1}R\Omega}{(1+\Gamma^{2}+2\Gamma R)}$;
$m_{24}=\kappa_{12}$; $m_{31}=\kappa_{21}$; $m_{32}=K+\frac{\beta_{22}}{2}\Omega^{2}-\kappa_{21}h^{-1}+\frac{\gamma_{2}Rh^{2}}{(1+\Gamma Rh^{2})}+\frac{\gamma_{2}s_{2}Rh^{2}\Omega(2+\Gamma Rh^{2})}{(1+\Gamma Rh^{2})^{2}}+\frac{i\gamma_{2}T_{R2}Rh^{2}\Omega}{(1+\Gamma^{2}+2\Gamma Rh^{2})}$;
$m_{33}=0$; $m_{34}=\frac{\gamma_{2}Rh^{2}}{(1+\Gamma Rh^{2})^{2}}$;
$m_{41}=-K+\frac{\beta_{21}}{2}\Omega^{2}-\kappa_{12}h+\frac{\gamma_{1}R}{(1+\Gamma R)}+\frac{\gamma_{1}s_{1}R\Omega(2+\Gamma R)}{(1+\Gamma R)^{2}}+\frac{i\gamma_{1}T_{R1}R\Omega}{(1+\Gamma^{2}+2\Gamma R)}$;
$m_{42}=\kappa_{12}$; $m_{43}=\frac{\gamma_{1}R}{(1+\Gamma R)^{2}}$;
and $m_{44}=0.$

The determinant of the matrix $M$ leads to a fourth order polynomial
in $K$, where the roots should possess a nonzero and negative imaginary
part that corresponds to a dispersion relation. MI occurs when the
wave number possesses a nonzero imaginary part leading to an exponential
growth of the perturbed amplitude. The MI is measured by power gain,
and it is defined at any pump frequency as \cite{Tatsing_JOSAB12}
\begin{equation}
G(\Omega)\equiv|\Im\left\{ K_{\max}\right\} |,\label{gain}
\end{equation}
where $\Im\left\{ K_{\max}\right\} $ denotes the imaginary part of
the root with the largest value $K_{\max}(\Omega)$.

\begin{figure*}[tb]
\includegraphics[width=0.48\columnwidth]{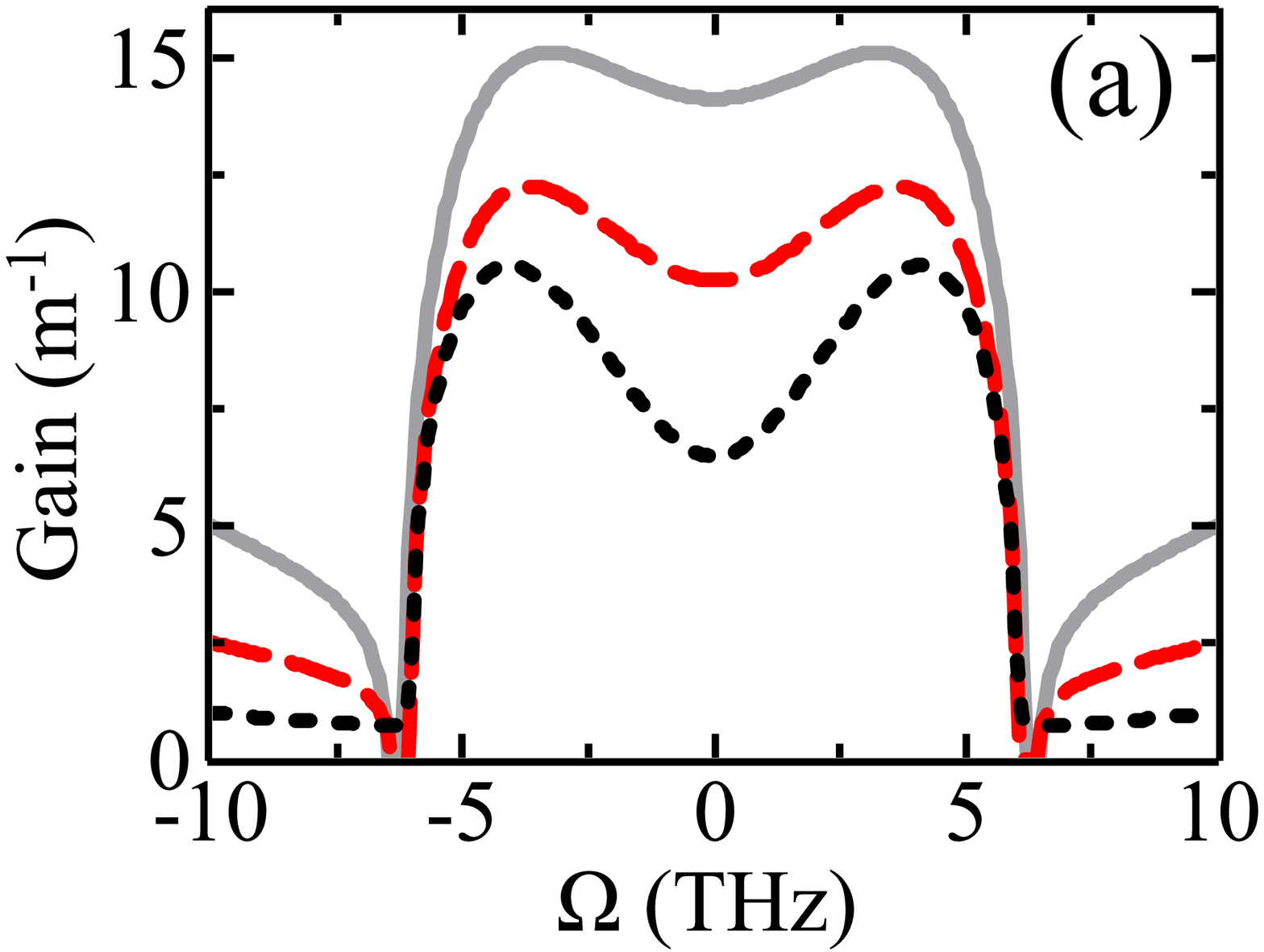} \includegraphics[width=0.48\columnwidth]{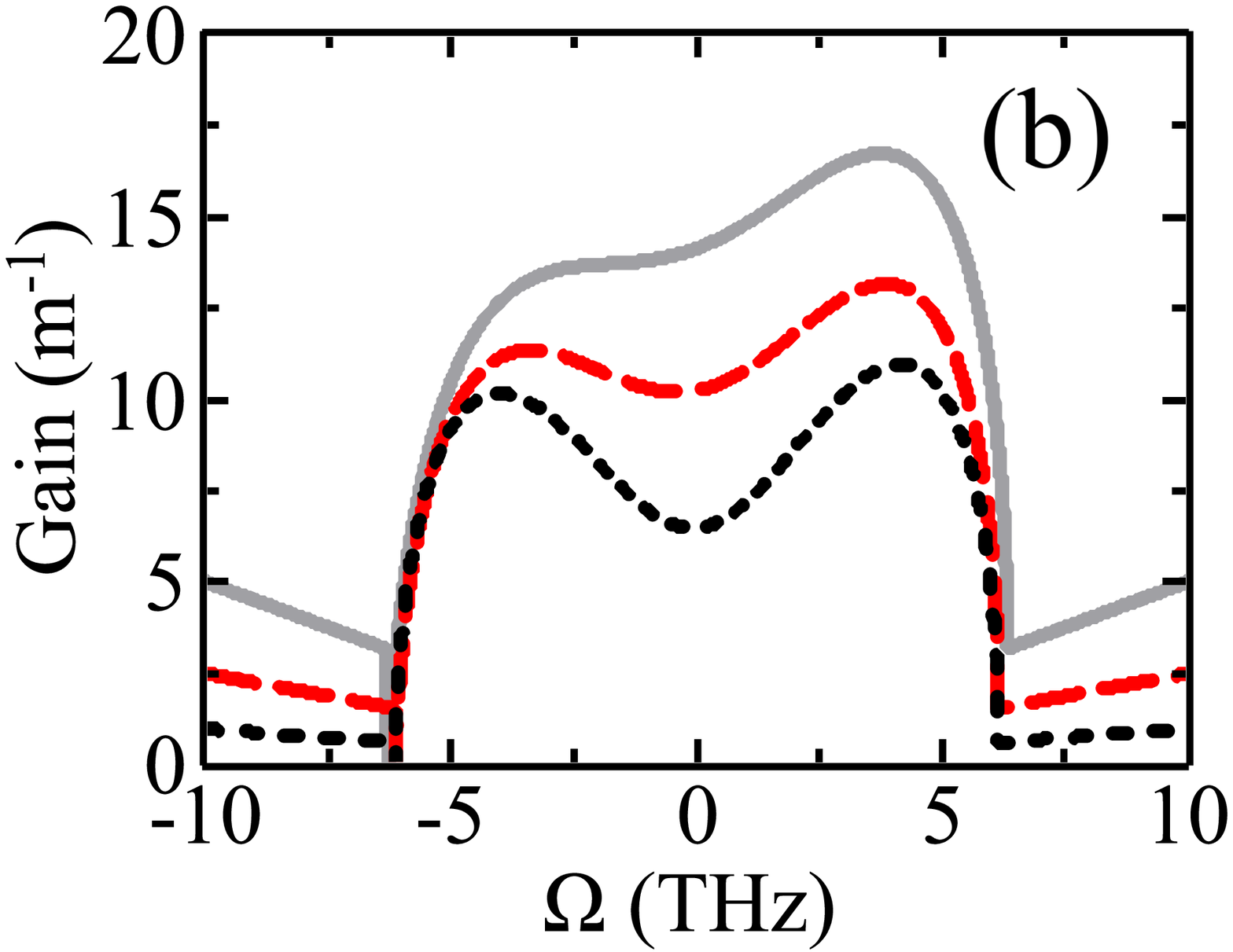}
\includegraphics[width=0.48\columnwidth]{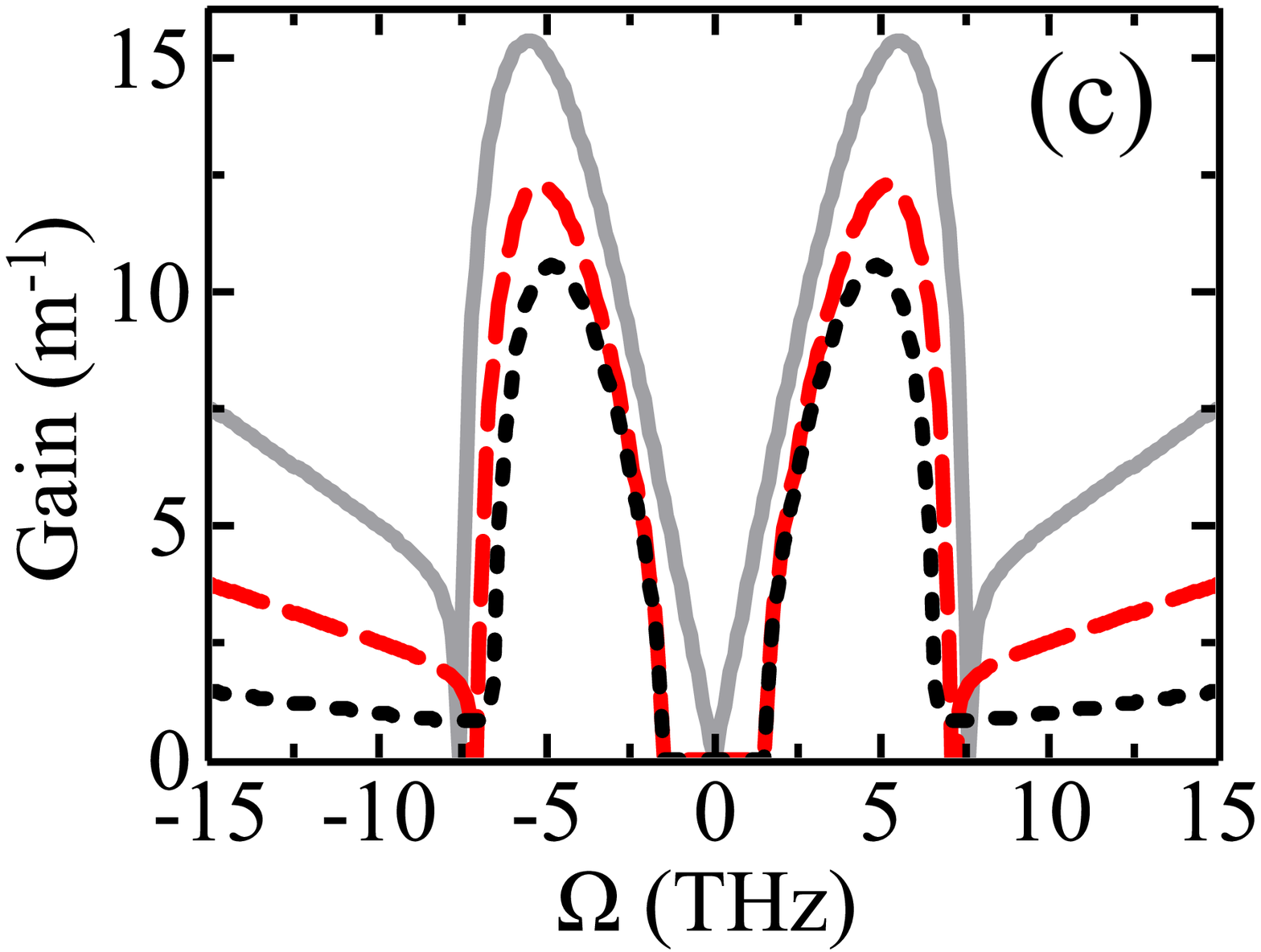} \includegraphics[width=0.48\columnwidth]{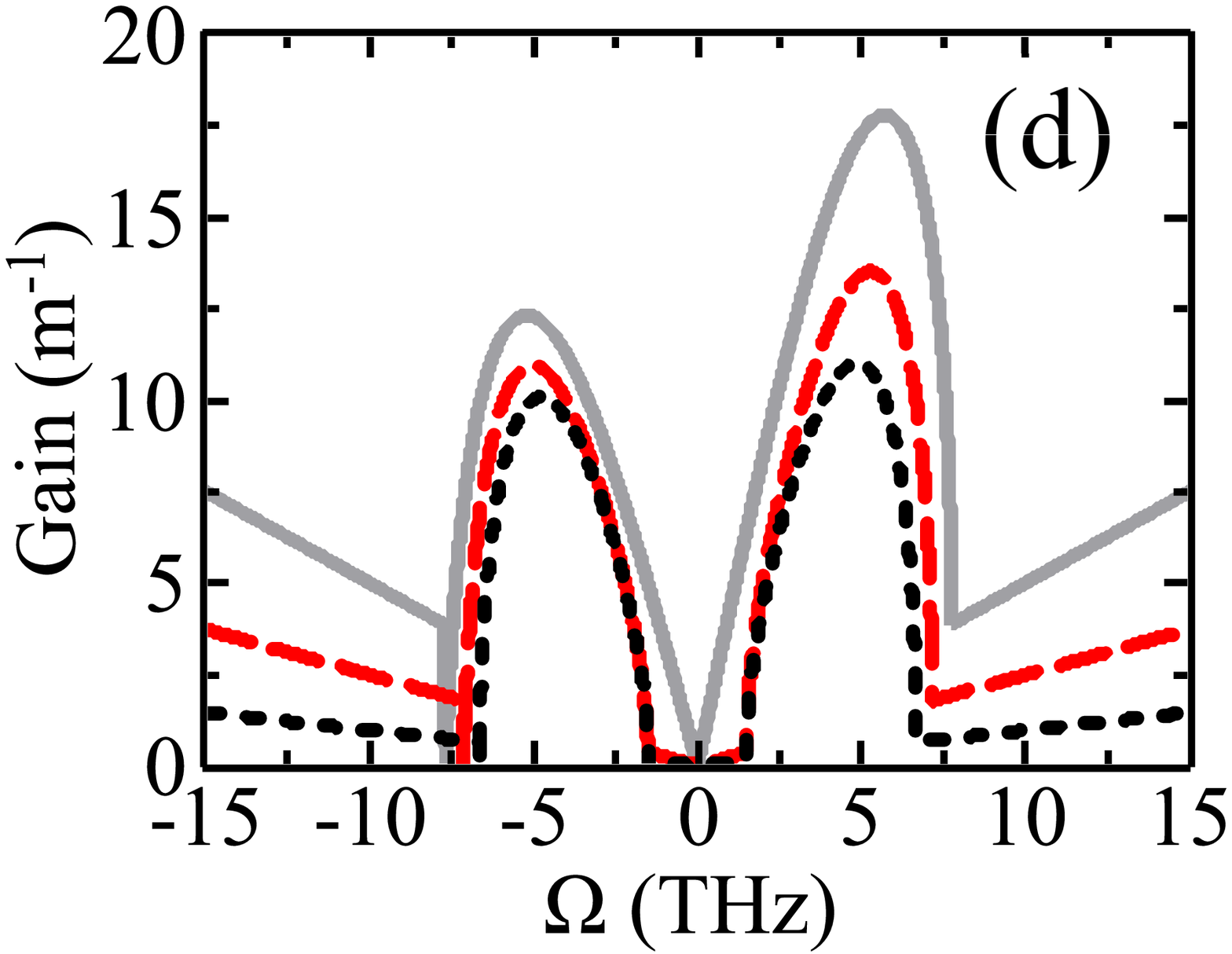}

\protect\caption{(Color online) Instability gain spectra showing the effect of intrapulse
Raman scattering in normal group velocity dispersion regime ($h=1$
and $\beta_{21}=\beta_{22}=1\:\mathrm{ps^{2}\,m^{-1}}$) for (a) $T_{R1}=T_{R2}=0.1\:\mathrm{ps/(kW\,m)}$
and (b) $T_{R1}=-T_{R2}=0.1\:\mathrm{ps/(kW\,m)}$ and in anomalous
group velocity dispersion regime ($h=-1$ and $\beta_{21}=\beta_{22}=-1\:\mathrm{ps^{2}\,m^{-1}}$)
for (c) $T_{R1}=T_{R2}=0.1\:\mathrm{ps/(kW\,m)}$ and (d) $T_{R1}=-T_{R2}=0.1\:\mathrm{ps/(kW\,m)}$.
The other parameters are $P=10\:\mathrm{kW}$, $\gamma_{1}=\gamma_{2}=1/(\mathrm{kW\,m})$,
$\kappa_{12}=\kappa_{21}=10\:\mathrm{m^{-1}}$, and $s_{1}=s_{2}=0$.
The saturation parameter values used herein are: $\Gamma=0$ in solid-line
(gray), $\Gamma=0.1\:\mathrm{kW^{-1}}$ in dashed-line (red), and
$\Gamma=0.4\:\mathrm{kW^{-1}}$ in dotted-line (black).}

\label{F5}
\end{figure*}

\begin{figure*}[tb]
\includegraphics[width=0.65\columnwidth]{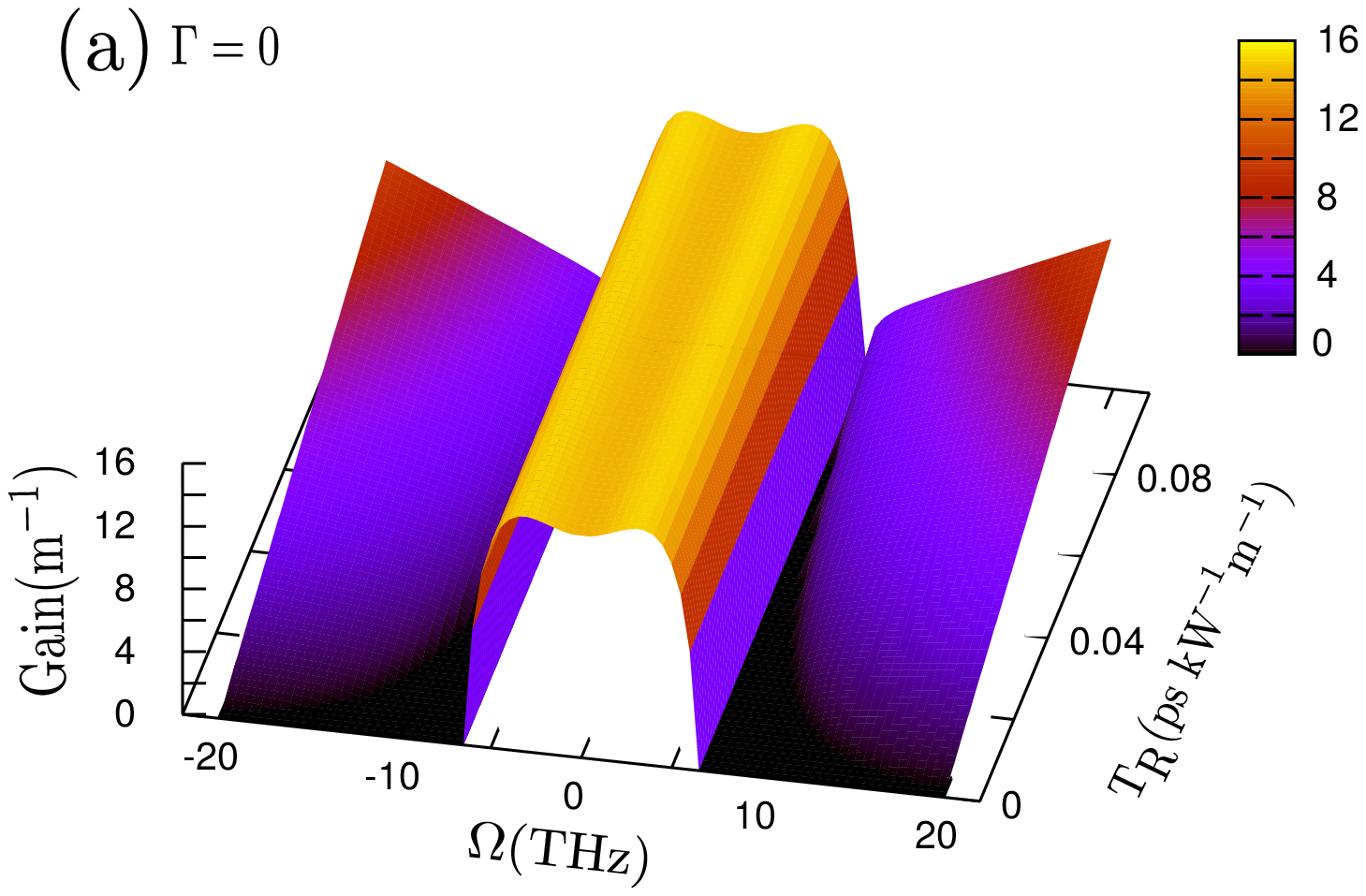} \includegraphics[width=0.65\columnwidth]{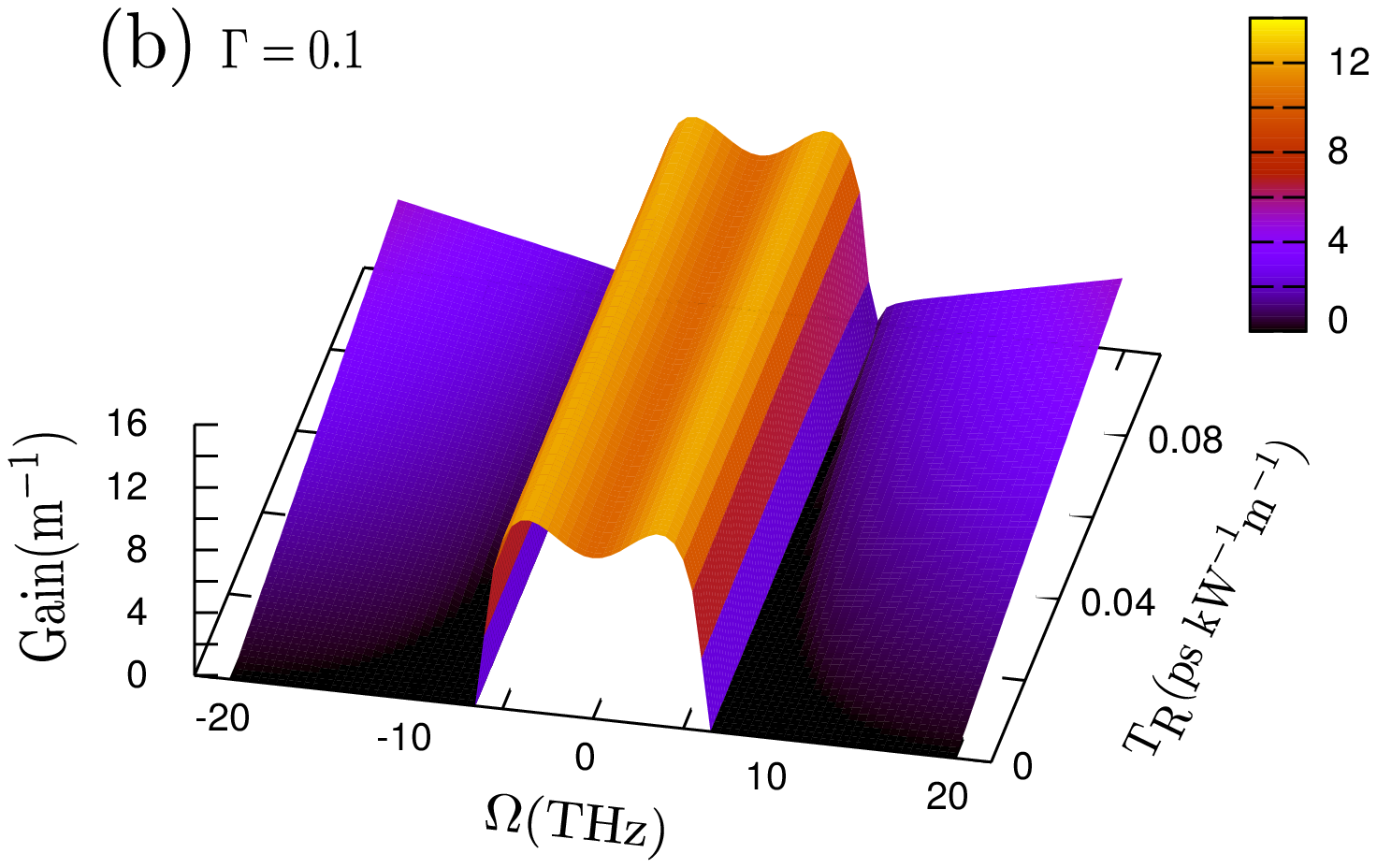}
\includegraphics[width=0.65\columnwidth]{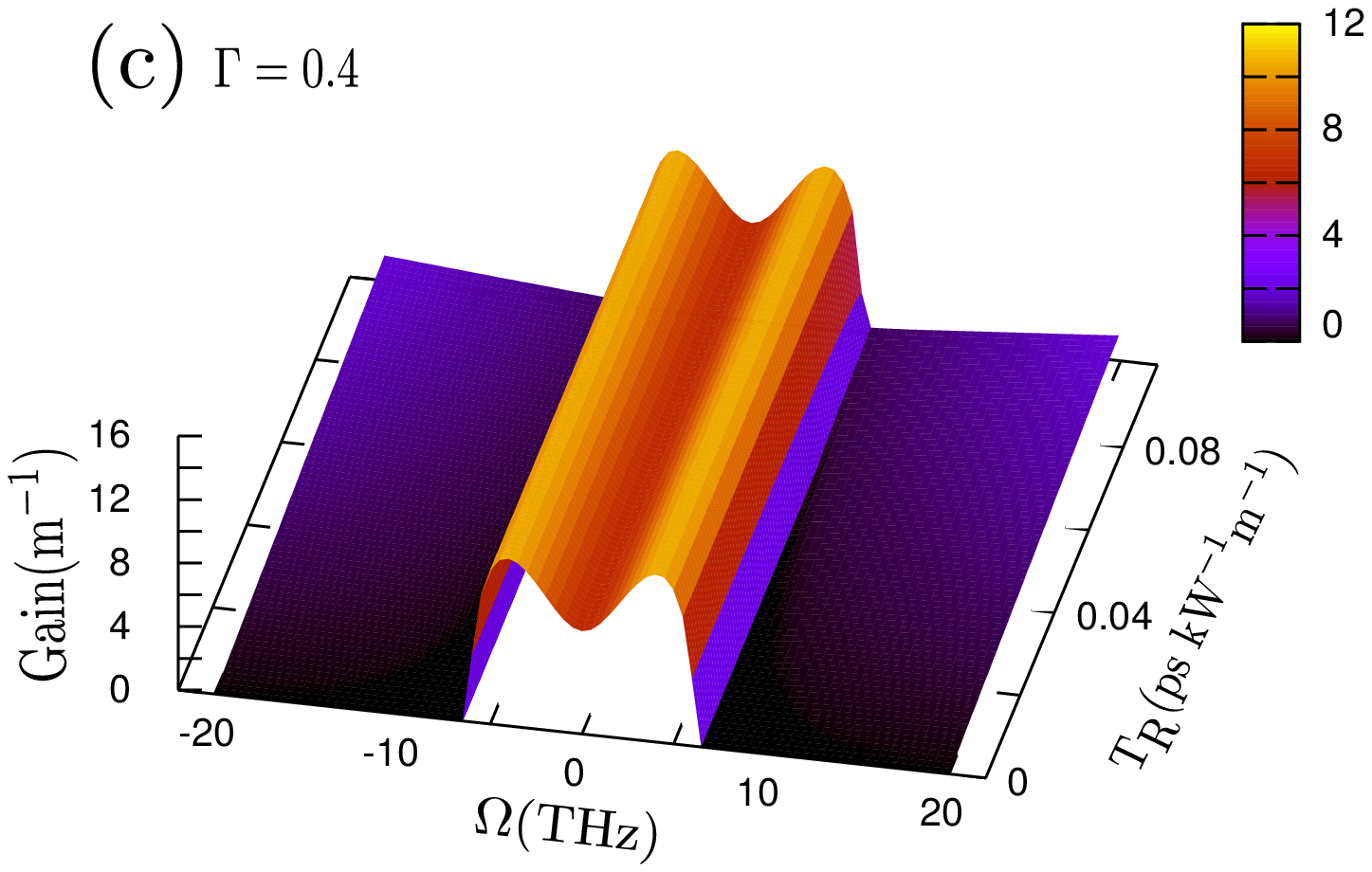}

\protect\caption{(Color online) Instability gain spectra versus the effect of intrapulse
Raman scattering (with $T_{R1}=T_{R2}=T_{R}$) in normal group velocity
dispersion regime for different values of saturation parameter. The
other parameters are the same ones used in Fig. \ref{F5}(a).}

\label{F6}
\end{figure*}

\begin{figure*}[tb]
\includegraphics[width=0.65\columnwidth]{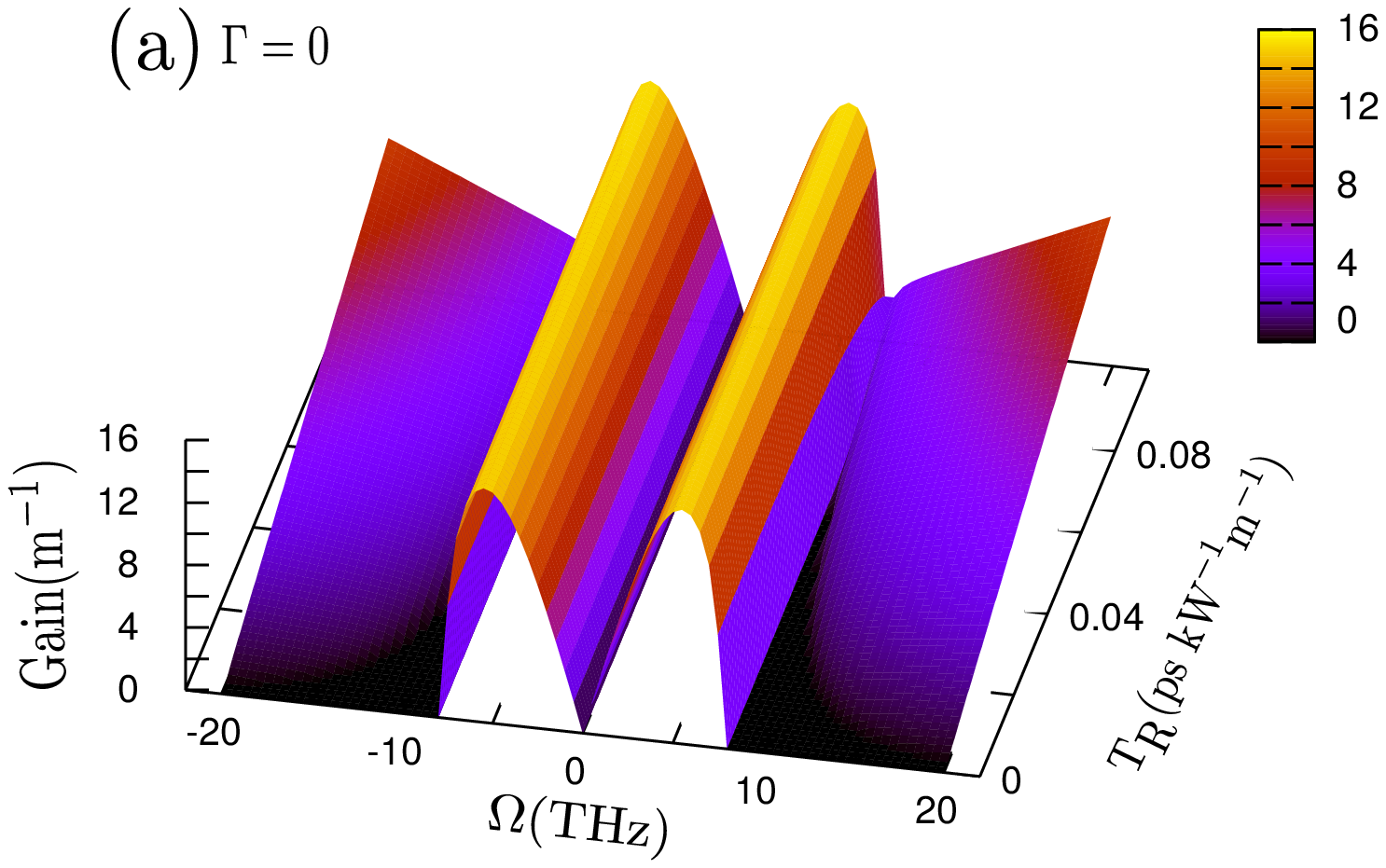} \includegraphics[width=0.65\columnwidth]{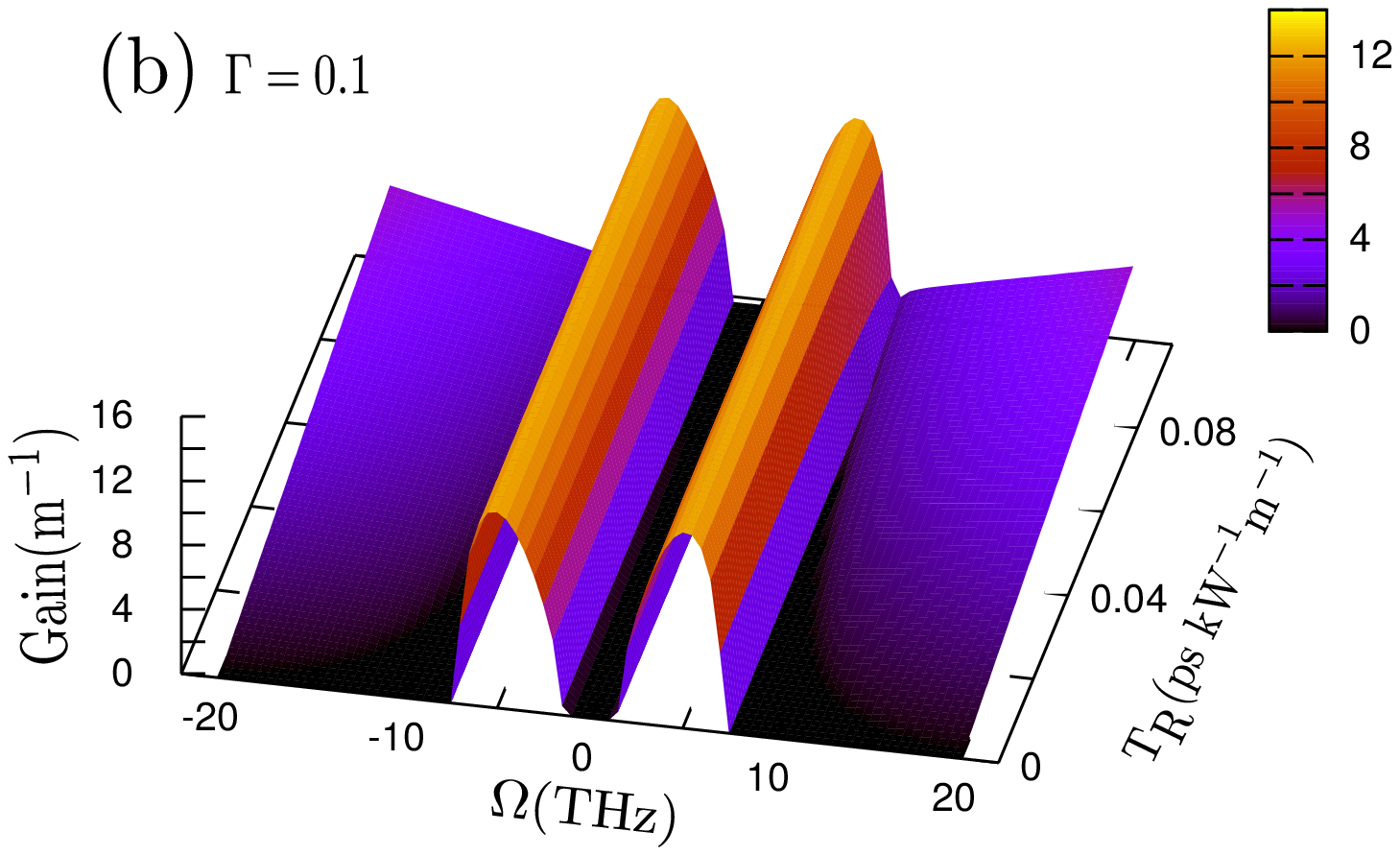}
\includegraphics[width=0.65\columnwidth]{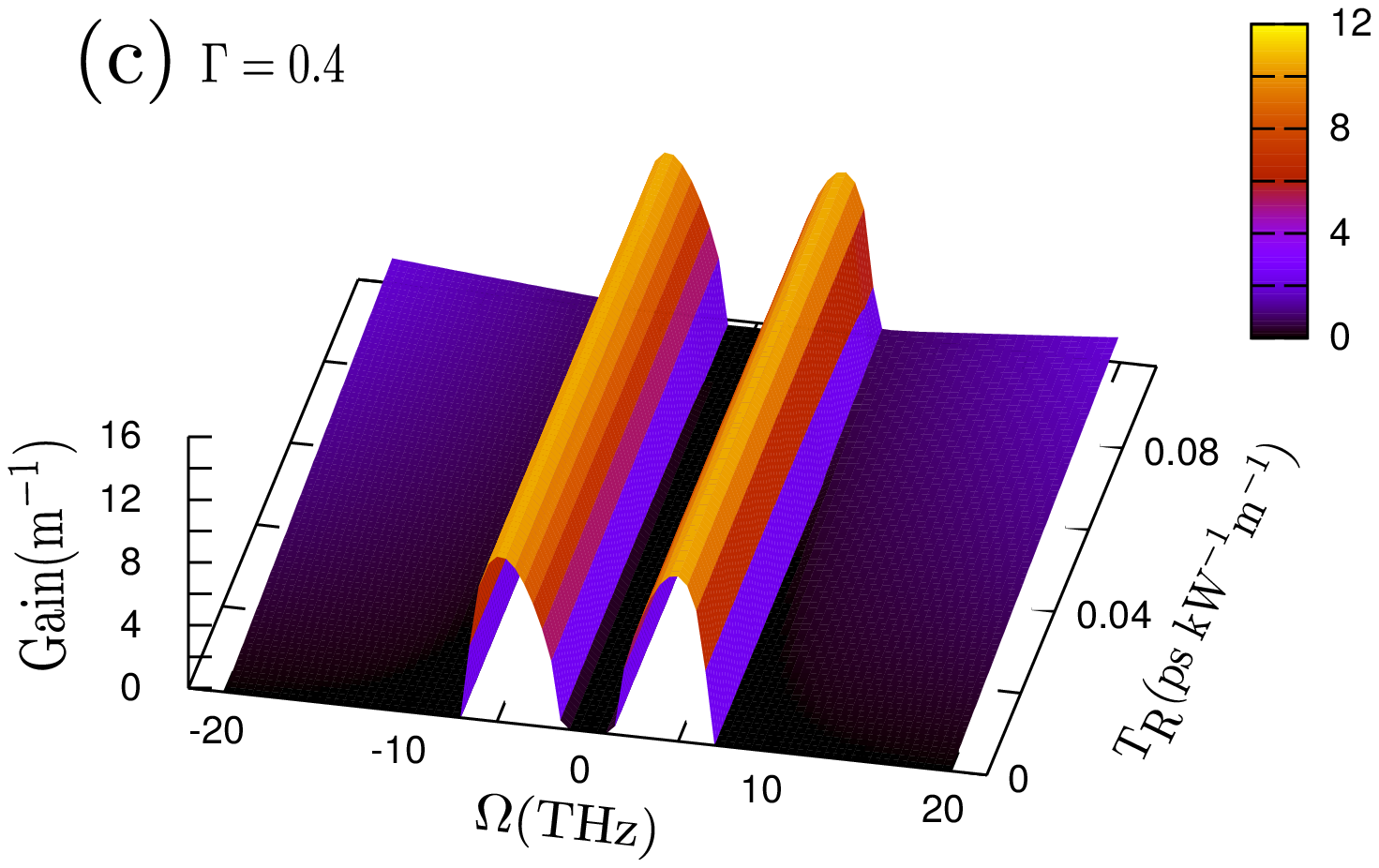}

\protect\caption{(Color online) Instability gain spectra versus the effect of intrapulse
Raman scattering (with $T_{R1}=T_{R2}=T_{R}$) in anomalous group
velocity dispersion regime for different values of saturation parameter.
The other parameters are the same ones used in Fig. \ref{F5}(c).}

\label{F7}
\end{figure*}

\section{Numerical results\label{sec:Numerical-results}}

\subsection{Effect of self-steepening on modulation instability\label{sub:Effect-of-self-steepening}}

Firstly we focus on the influence of self-steepening effect on MI
in oppositely directed coupler for different values of saturation
parameter. To this end, we omit the Raman self-scattering effect by
setting $T_{R1}=T_{R2}=0$. In Fig. \ref{F1} we consider MI in normal
group velocity in dispersion regime under different combinations of
$s_{1}$ and $s_{2}$, considering $h=1$ and $\beta_{21}=\beta_{22}=1\:\mathrm{ps^{2}\,m^{-1}}$,
for simplicity. Also, the MI for the anomalous case of group velocity
dispersion is illustrated in Fig. \ref{F2}, where we have used $h=-1$
and $\beta_{21}=\beta_{22}=-1\:\mathrm{ps^{2}\,m^{-1}}$. The other
parameters that we used were $\kappa_{12}=\kappa_{21}=10\:\mathrm{m}^{-1}$,
$\gamma_{1}=\gamma_{2}=1/(\mathrm{kW\,m})$, and $P=10\:\mathrm{kW}$. 

\emph{Normal group velocity dispersion} - To be more specific, in
Fig. \ref{F1}(a) we display the case without self-steepening effects
($s_{1}=s_{2}=0$) for three different values of saturation parameter:
$\Gamma=0$ in solid-line (gray), $\Gamma=0.1\:\mathrm{kW^{-1}}$
in dashed-line (red), and $\Gamma=0.4\:\mathrm{kW^{-1}}$ in dotted-line
(black). These parameters are also used in Figs. \ref{F1}(b)-(d).
Note in Fig. \ref{F1}(a) that the instability spectra consists of
single conventional MI band centered at zero perturbation frequency
formed by balance between group velocity dispersion and self-phase
modulation. However, when increasing the value of the saturation parameter
the instability gain decreases faster for frequencies near to $\Omega=0$.
We stress that in the present model, for the normal group velocity
dispersion regime, the instability gain exists even if perturbation
frequency is zero \cite{Ali_PRE14}.

In Fig. \ref{F1}(b) we study the self-steepening effect in only one
of the channels by tuning $s_{1}=0$ and $s_{2}=1\:\mathrm{ps}/(\mathrm{kW\,m})$.
In this figure is evident the influence of the value of the saturation
parameter on the number of MI bands. The greater the value of this
parameter is, more centralized will be the gain region, thereby decreasing
the amount of MI bands. The results for the two channels with the
self-steepening effect are shown in Fig. \ref{F1}(c), where we have
used $s_{1}=s_{2}=1\:\mathrm{ps}/(\mathrm{kW\,m})$. Note that without
saturation ($\Gamma=0$) there are two MI bands centered close to
$\Omega=\pm20\:\mathrm{THz}$ plus three close to $\Omega=0$. By
increasing the saturation parameter for $\Gamma=0.1\:\mathrm{kW^{-1}}$
we observe only three bands centered close to $\Omega=0$. And, when
setting $\Gamma=0.4\:\mathrm{kW^{-1}}$ we see the appearance of only
one MI band. This is also evident when one looks at Fig. \ref{F3},
where we show a 3D surface of the instability gain as a function of
the perturbation frequency and self-steepening parameter $s_{1}$,
for three different values of the saturation parameter, i.e., $\Gamma=0$
in Fig. \ref{F3}(a), $\Gamma=0.1\:\mathrm{kW^{-1}}$ in Fig. \ref{F3}(b),
and $\Gamma=0.4\:\mathrm{kW^{-1}}$ in Fig. \ref{F3}(c).

Fig. \ref{F1}(d) displays the case in which both channels are influenced
by self-steepening effect but opposite in sign (here we set $s_{1}=-s_{2}=1\:\mathrm{ps}/(\mathrm{kW\,m})$).
Note that, in Fig. \ref{F1}(d), as well as shown in Ref. \cite{Ali_PRE14}
for the case without saturation, the gain presents a similar behavior
to the case where both channels are not influenced by self-steepening
effect (Fig. \ref{F1}(a)). Indeed, we have observed that the effective
influence of self-steepening parameters depends on its algebraic sum,
i.e., this effect is canceled in the gain shown by Fig. \ref{F1}(d).
This can be checked by comparing the dashed or dotted-lines of Figs.
\ref{F1}(a) and \ref{F1}(d).

\emph{Anomalous group velocity dispersion} - Differently from the
normal dispersion, in the anomalous case there is no instability gain
at zero perturbation frequency. Also, as previously verified \cite{Ali_PRE14},
one can see in Fig. \ref{F2} that the maximum gain and the band width
are influenced by the presence of self-steepening effect. However,
when saturation is present one can verify a gain reduction and a separation
of the bands. This is evident when one look at the 3D surfaces of
the instability gain for different values of the saturation parameter:
$\Gamma=0$ in Fig. \ref{F4}(a), $\Gamma=0.1\:\mathrm{kW^{-1}}$
in Fig. \ref{F4}(b), and $\Gamma=0.4\:\mathrm{kW^{-1}}$ in Fig.
\ref{F4}(c).

In Fig. \ref{F2}(a) we display the case without self-steepening effects
($s_{1}=s_{2}=0$) for three different values of saturation parameter:
$\Gamma=0$ in solid-line (gray), $\Gamma=0.1\:\mathrm{kW^{-1}}$
in dashed-line (red), and $\Gamma=0.4\:\mathrm{kW^{-1}}$ in dotted-line
(black). Note the differences we just mentioned comparing the Fig.
\ref{F2}(a) with Fig. \ref{F1}(a). Also, differently from the single
null point that we obtain at $\Omega=0$ for the case without saturation,
in the presence of saturation we observe a large hole in the gain
region near $\Omega=0$.

In Fig. \ref{F2}(b) we show the self-steepening effect in only one
of the channels by tuning $s_{1}=0$ and $s_{2}=1\:\mathrm{ps}/(\mathrm{kW\,m})$.
By increasing the value of the saturation parameter we observe a compactification
of instability bands, a result similar to that obtained for normal
dispersion, but now with null gain for disturbance near zero. This
can also be seen in Fig. \ref{F2}(c), in which we consider two channels
with the self-steepening effect (by adjusting $s_{1}=s_{2}=1\:\mathrm{ps}/(\mathrm{kW\,m})$).

The case in which both channels are influenced by self-steepening
effect but opposite in sign (considered in Fig. \ref{F2}(d) for $s_{1}=-s_{2}=1\:\mathrm{ps}/(\mathrm{kW\,m})$)
shows the same behavior when compared to the case without self-steepening
effects (Fig. \ref{F2}(a)).

\subsection{Effect of intrapulse Ramam scattering on modulation instability\label{sub:Effect-of-intrapulse}}

Next, we study the effect of intrapulse Raman scattering on MI in
oppositely directed couplers as a function of saturation parameter.
For this particular study, we neglect the role of self-steepening
effect (by setting $s_{1}=s_{2}=0$). Here, we use the same parameters
used in the previous subsection, i.e., $\kappa_{12}=\kappa_{21}=10\:\mathrm{m}^{-1}$,
$\gamma_{1}=\gamma_{2}=1/(\mathrm{kW\,m})$, and $P=10\:\mathrm{kW}$.

In Figs. \ref{F5}(a) and \ref{F5}(b) we consider MI in normal group
velocity in dispersion regime under an intrapulse Raman scattering
effect with amplitude $T_{R1}=T_{R2}=0.1\:\mathrm{ps/(kW\,m)}$ and
$T_{R1}=-T_{R2}=0.1\:\mathrm{ps/(kW\,m)}$, respectively, and considering
$h=1$ and $\beta_{21}=\beta_{22}=1\:\mathrm{ps^{2}\,m^{-1}}$, for
simplicity. Also, the MI for the anomalous case of group velocity
dispersion is illustrated in Figs. \ref{F5}(c) and \ref{F5}(d),
where we have $T_{R1}=T_{R2}=0.1\:\mathrm{ps/(kW\,m)}$ and $T_{R1}=-T_{R2}=0.1\:\mathrm{ps/(kW\,m)}$,
respectively, plus $h=-1$ and $\beta_{21}=\beta_{22}=-1\:\mathrm{ps^{2}\,m^{-1}}$.
The solid-lines (gray) in Figs. \ref{F5}(a)-(d) represent the cases
without saturation effect ($\Gamma=0$) while in the dashed-lines
(red) and dotted-lines (black) we analyzed the effects of the saturation
characterized by the parameters $\Gamma=0.1\:\mathrm{kW^{-1}}$ and
$\Gamma=0.4\:\mathrm{kW^{-1}}$, respectively. Note that in the absence
of intrapulse Raman scattering, one retrieves the results shown in
Figs. \ref{F1}(a) and \ref{F2}(a) for the cases of normal and anomalous
dispersion, respectively. Then, by comparing the Figs. \ref{F5}(a)-(d)
with Figs. \ref{F1}(a) and \ref{F2}(a) one observes the appearance
of a tail for highest perturbation frequencies that makes de instability
gain to increase linearly in these regions. However, by increasing
the saturation effect the maximum in these instability gain are reduced
also in its tails (see dashed-lines and dotted lines in Fig. \ref{F5}).
Also, differently from the cases shown in Figs. \ref{F5}(a) and \ref{F5}(c),
when the Raman scattering coefficients are opposite in sign there
will be a symmetry breaking in the gain regions (Figs. \ref{F5}(b)
and \ref{F5}(d)).

Finally, by increasing the Raman scattering intrapulse will cause
a linear increase in the maximum MI in the tail regions while maintaining
the core bands largely unchanged. This can be visualized in the 3D
surfaces shown in Figs. \ref{F6} and \ref{F7} for the cases of normal
and anomalous dispersion, respectively, with $T_{R1}=T_{R2}=T_{R}$
and for three values of the saturation parameter.

\section{Conclusion\label{sec:Conclusion}}

To summarize we have studied the modulation instability in oppositely
directed coupler with higher-order effects and saturable nonlinearities
for both normal and anomalous group velocity dispersion regimes. It
is found that the instability bands under saturation effects, besides
reducing its amplitude, it also makes the bands gathered together
around the zero perturbation frequency when the system presents self-steepening
effects. However, in anomalous group velocity dispersion regime this
compactification due to the growth of the saturation parameter is
followed by an increase in the null gain region near $\Omega=0$,
enlarging the separation of the instability bands close to this point. 

In the case of intrapulse Raman scattering, new instability regions
are created in the tail of the original MI bands (by original bands
we emphasize the case without this effect). In this case the saturation
effects causes a change in the center of the gain region decreasing
the width of the bands, followed by an attenuation in the amplitude
of the instability gain. Also, the symmetry shown in the MI regions,
when the Raman scattering coefficients are equal in sign, are broken
for Raman scattering coefficients opposite in sign. Finally, the present
study reinforces those results presented in Ref. \cite{Ali_PRE14}
providing a new way to generate solitons or ultrashort pulses in oppositely
directed coupler with saturable nonlinearities.
\begin{acknowledgments}
We thank the Brazilian agencies CNPq, CAPES, FAPEG, and Instituto
Nacional de Ci\^encia e Tecnologia-Informaç\~ao Qu\^antica (INCT-IQ) for
partial support.\end{acknowledgments}

\end{document}